\documentclass[letterpaper,twocolumn,10pt]{article}
\usepackage{usenix2021}

\usepackage[most]{tcolorbox}
\newtcolorbox[auto counter,number within=section]{AlgorithmBox}[2][]
{
  title    = Algorithm~\thetcbcounter~{#2},{#1},
  fonttitle=\bfseries,
  enforce breakable, 
  colframe = black,
  coltitle = black,
  colbacktitle = white, 
  coltext  = black, 
  colback  = white,
  arc=0pt, 
  outer arc=0pt,
  boxrule=1pt,
  left=2pt, 
  right=2pt, 
  top = 0pt,
  bottom = 0pt,
  leftrule=0mm,
  rightrule=0mm
}

\makeatletter
\newcommand{\@chapapp}{\relax}%
\makeatother

\usepackage{xparse}

\let\labelindent\relax
\usepackage{enumitem}
\usepackage{balance}
\usepackage{subfigure}
\usepackage{epsfig}
\usepackage{graphicx}
\usepackage{amsmath}
\usepackage{amssymb}
\usepackage{amsfonts}
\usepackage{verbatim}
\usepackage{graphicx,url}
\usepackage{algorithm} 
\usepackage{stackrel}
\usepackage{mathtools}
\usepackage[noend]{algpseudocode}
\usepackage{multicol}
\usepackage{float}
\usepackage{color}
\usepackage{booktabs}
\usepackage{wrapfig}
\usepackage{filecontents}
\usepackage{xspace}
\usepackage{framed}
\usepackage{setspace}
\usepackage{thmtools, thm-restate}
\usepackage{etoolbox, environ}
\usepackage{makecell}
\usepackage{xcolor}

\usepackage{amsthm, amsmath} 
\usepackage[normalem]{ulem}
\usepackage{pgfplots}

\usepackage{comment}

\algnewcommand{\LineComment}[1]{\Statex \(//\) #1}
\newcommand{\ceil}[1]{\left\lceil #1 \right\rceil}

{\hfil\hskip2em\penalty250\parfillskip=0pt\finalhyphendemerits=0$\qed$%
\par\smallskip\par}

\usetikzlibrary{matrix,shapes,arrows,positioning,chains, calc}

   {\endMakeFramed}

\tikzstyle{every picture}+=[remember picture]
\tikzstyle{na} = [shape=rectangle,inner sep=0pt,text depth=0pt]
\usepackage{amssymb}

\usepackage[font=small,skip=0pt]{caption}
\usepackage{pifont}
\newcommand{\cmark}{\ding{51}}%
\newcommand{\xmark}{\ding{55}}%

\newcommand{\mytodo}[4]{%
\ifx&#3&\textcolor{#1}{#4}%
\else\tikz\node[na](dst){};%
\marginpar{\raggedright\footnotesize\textcolor{#1}{%
$\lll$#2$\ggg$%
\tikz\node[na](src){};\\
#3}}%
\textcolor{#1}{#4}%
\begin{tikzpicture}[overlay]
  \path[->,#1,thin,dashed](src) edge [out=0] (dst);
\end{tikzpicture}%
\fi}

\newcommand{\sys}{Wink}

\definecolor{bblue}{HTML}{4F81BD}
\definecolor{rred}{HTML}{C0504D}
\definecolor{ggreen}{HTML}{9BBB59}
\definecolor{ppurple}{HTML}{9F4C7C}

\newenvironment{myquote}%
  {\smallskip \list{}{\leftmargin=0.2in\rightmargin=0.2in}\item[] \em}%
  {\endlist}

\newtheorem{theorem}{Theorem}
\renewcommand{\paragraph}[1]{\noindent {\bf #1.}~} 
\newtheorem{defn}{Definition}

\newenvironment{proofsketch}{\par\noindent\textsl{Proof (sketch):}\kern1ex}%
{\hfil\hskip2em\penalty250\parfillskip=0pt\finalhyphendemerits=0$\qed$%
	\par\smallskip\par}

\newcommand{\BT}{\begin{theorem}}
\newcommand{\ET}{\end{theorem}}

\newcommand{\BD}{\begin{definition}}
\newcommand{\ED}{\end{definition}}

\newcommand{\BCR}{\begin{corollary}}
\newcommand{\ECR}{\end{corollary}}

\newcommand{\BEX}{\begin{example}}
\newcommand{\EEX}{\end{example}}

\newcommand{\BL}{\begin{lemma}}
\newcommand{\EL}{\end{lemma}}

\newcommand{\BP}{\begin{proposition}}
\newcommand{\EP}{\end{proposition}}

\newcommand{\BCM}{\begin{claim}}
\newcommand{\ECM}{\end{claim}}

\newcommand{\BPF}{\begin{proof}}
\newcommand{\EPF}{\end{proof}}
\newcommand{\BEN}{\begin{enumerate}}
\newcommand{\EEN}{\end{enumerate}}

\newcommand{\BI}{\begin{itemize}}
\newcommand{\EI}{\end{itemize}}

\newcommand{\BO}{\begin{observation}}
\newcommand{\EO}{\end{observation}}

\newcommand{\BDS}{\begin{description}}
\newcommand{\EDS}{\end{description}}

\newcommand{\getsr}{\gets_{\mbox{\tiny R}}}

\def\squareforqed{\hbox{\(\blacksquare\)}}
\def\qed{\hfill \squareforqed}

\newcommand{\etal}{{\it et al }}

\newcommand{\ignore}[1]{}

\def\encrypt{\text{\sf{Encrypt}}\xspace}
\def\decrypt{\text{\sf{Decrypt}}\xspace}

\newcounter{defcounter}
\newlength{\protowidth}

\newcommand{\adv}{\ensuremath{\mathcal{A}}\xspace}
\newcommand{\ch}{\ensuremath{\mathcal{C}}\xspace}

\newcommand{\olrk}[1]{%
   \ifx\nursymbol#1\else\!\!\mskip4.5mu plus 0.5mu\left(#1\right)\fi}
\newcommand{\elrk}[1]{%
   \ifx\nursymbol#1\else%
        \!\!\mskip4.5mu plus0.5mu\left[\mskip2.5mu plus0.5mu #1\right]\fi}

\newcommand{\tz}{TrustZone}

\newcommand{\registration}{\ensuremath{\mathsf{Registration}}\xspace}
\newcommand{\keyDerive}{\ensuremath{\mathsf{KeyDerive}}\xspace}
\newcommand{\sendMessage}{\ensuremath{\mathsf{SendMsg}}\xspace}
\newcommand{\recvMessage}{\ensuremath{\mathsf{RecvMsg}}\xspace}
\newcommand{\secParam}{\ensuremath{\lambda}\xspace}
\newcommand{\clientID}{\ensuremath{\mathsf{ID}}\xspace}
\newcommand{\keyBundle}{\ensuremath{\mathcal{K}}\xspace}
\newcommand{\genericMsg}{\ensuremath{\mathsf{msg}}\xspace}
\newcommand{\genericHidMsg}{\ensuremath{\mathsf{msg_H}}\xspace}
\newcommand{\encryptedMsg}{\ensuremath{c}\xspace}
\newcommand{\defeq}{\ensuremath{~\stackrel{\mathclap{\scriptsize\mbox{def}}}{=}~}\xspace}
\newcommand{\genericRandCoin}{\ensuremath{r}\xspace}
\newcommand{\genericEncryptionSchemeDef}{\ensuremath{\mathcal{S} \defeq \{\mathsf{KeyGen}, \encrypt, \decrypt\}}\xspace}
\newcommand{\genericEToEE}{\ensuremath{\mathcal{M}}\xspace}
\newcommand{\winkEToEE}{\ensuremath{\mathcal{W}}\xspace}
\newcommand{\hidKeyDerive}{\ensuremath{\mathsf{HidKeyDerive}}\xspace}
\newcommand{\sendHidMessage}{\ensuremath{\mathsf{SendHidMsg}}\xspace}
\newcommand{\recvHidMessage}{\ensuremath{\mathsf{RecvHidMSg}}\xspace}
\newcommand{\hidKey}{\ensuremath{\kappa_H} \xspace}
\newcommand{\genericEncryptionScheme}{\ensuremath{\mathcal{S}}\xspace}
\newcommand{\genericBit}{\ensuremath{b}\xspace}
\newcommand{\advBit}{\ensuremath{b'}\xspace}
\newcommand{\prob}[1]{\ensuremath{\mathbb{P}\left [#1 \right]}\xspace}

\renewcommand{\getsr}{\ensuremath{\stackrel{\$}{\leftarrow}}\xspace}

\begin{document}

\title{{\sys}: Deniable Secure Messaging}

\author{
  {\rm Anrin Chakraborti $^*$}\\
  Duke University
  \and
  {\rm Darius Suciu $^*$}\\
    Stony Brook University
  \and
  {\rm Radu Sion}\\
  Stony Brook University
}
\maketitle
\def\thefootnote{*}\footnotetext{These authors contributed equally to this work}\def\thefootnote{\arabic{footnote}}


\maketitle

\begin{abstract}
End-to-end encrypted (E2EE) messaging is an essential first step in providing message confidentiality. Unfortunately, all security guarantees of end-to-end encryption are lost when keys or plaintext are disclosed, either due to device compromise or (sometimes lawful) coercion by powerful adversaries. This work introduces Wink, the first \textit{plausibly-deniable messaging} system protecting message confidentiality from partial device compromise and compelled key disclosure.  Wink can surreptitiously inject hidden messages in standard random coins (e.g., salts, IVs) used by existing E2EE protocols. It does so as part of legitimate secure cryptographic functionality deployed inside the widely-available trusted execution environment (TEE) TrustZone. This results in hidden communication using virtually unchanged {\em existing E2EE messaging apps}, as well as strong plausible deniability. {\sys} has been demonstrated with multiple existing E2EE applications (including Telegram and Signal) with minimal (external) instrumentation, negligible overheads, and crucially, without changing on-wire message formats.

\end{abstract}

\section{Introduction}
\label{intro}
Secure messaging with end-to-end encryption has become the standard mode of communication between trusted peers.  Almost all messaging apps today support \textit{end-to-end encryption} for ensuring message confidentiality in the presence of compromised or malicious intermediate nodes; the messages are available in plaintext only on the user devices and encrypted when in transit with user-chosen keys.

However, if the encryption keys are eventually leaked to an adversary, message confidentiality is automatically lost. Unintended disclosure of cryptographic keys is not uncommon: (i) typical commercially-available user devices are vulnerable to adversary-controlled malware including backdoors into widely-deployed secure apps~\cite{{apple_fbi,whatsapp_backdoor,us_law_backdoor}}, and ii) users can be coerced to hand over cryptographic information (e.g., keys)~\cite{youthjailedforpassword,ukencryptionlaws,forcedsearchatUSborder,humanrightsRussia}.



Achieving truly private communication between trusted parties requires stronger security guarantees  than what standard end-to-end encryption can offer. While existing solutions can solve some facets of the problem, they ultimately fall short of addressing the problem in its entirety. Some notable candidates include off-the-record messaging (OTR) \cite{borisov2004OTR}, deniable encryption~\cite{sahaiSTOC14,canetti97deniable,canetti2020fullybideniable}, ephemeral messages, network stegnography \cite{kaptchuk2021meteor}. Most of these techniques either fail when the end devices are also compromised by the adversary (in addition to the network monitoring), or/and are impractical for real-world applications (see Section \ref{sec:related} for more detailed comparisons). However, perhaps more importantly, due to their impracticality (and in some cases incompatibility with other applications), most of these techniques are not ubiquitously deployed; thus the mere presence of an application implementing these techniques can indicate to an adversary that the application is being specifically used to hide information.
In light of these observations, this paper asks

\begin{myquote}
Can we achieve end-to-end encrypted communication between trusted peers when the adversary can obtain access (e.g., through adversary-controlled malware) into the user device, can read on-wire transcripts, and can compel the users to reveal corresponding encryption keys (and other metadata)?
\end{myquote}

At first glance, it may appear that efforts to answer this question are futile as the adversary controls the software on the compromised device and can observe messages in the plain by monitoring I/O channels. In fact, even without any sophisticated mechanisms, if the adversary is aware that a message has been sent/received (e.g., by monitoring memory usage, network I/O), then breaching confidentiality by compelling the user to hand over the keys is trivial. Thus, against such a strong adversary the only recourse is to not only hide the message contents but also the fact that a message has been exchanged. 
Unfortunately, without a safe haven for running some critical parts of software outside the adversary's control (and knowledge), and storing keys for end-to-end encryption, the task is perhaps impossible. 

This work answers the question affirmatively under a more realistic and wide-spread threat model, where the adversary only partially compromises the device, and shows that a safe haven (ensuring message confidentiality against the adversary) is indeed realizable with the help of a TEE~\footnote{Of course, if the adversary is capable of completely compromising a device even if it has advanced hardware-based defenses such as TEE isolation, no safe haven exists. Breaking TEE isolation is however significantly more difficult than user coercion and partial compromise through spyware, etc.}. However, the task is not without challenges. First, a TEE is not designed to provide plausible deniability, and so while the TEE may provide confidentiality of information, it will not hide the artifacts of the execution (e.g., the fact that a message has been sent/received). Second, messaging applications are complex, and executing all its constituent tasks inside a TEE is impractical; it also makes the other software in the TEE vulnerable due to the increased TCB. Finally, building an application for a TEE with the sole purpose of hiding message communication \textit{without any other plausible use-case} is not enough since the possession of such software renders the user a suspect.  

This paper presents the design and implementation of {\sys}, a \textit{plausibly-deniable messaging} framework, which addresses all of these challenges. Informally, {\em plausibly-deniable messaging} (PDM) enables trusted peers to exchange messages but later plausibly deny the exchange to an adversary by providing an alternate, plausible explanation for all the actions observed by the adversary (which includes network traffic and system artifacts). {\sys} realizes a safe haven by ensuring that the TEE is used minimally (with small TCB) and correctly (with secure communication channels to components outside the TEE). {\sys} is demonstrably compatible with two widely used E2EE messaging apps, namely Signal and Telegram, with only minimal instrumentation. Importantly, this integration does not alter the on-wire message format native to the messaging app(s).


\paragraph{{\sys} Cryptographic Library}
{\sys} is a library that performs cryptographic operations at the request of E2EE messaging applications. This includes generating and storing keys, random coins, etc. and encrypting/decrypting messages. {\sys} runs inside the TEE and integrates 
with a standard E2EE messaging application that runs outside the TEE. In this way, only the cryptographic operations for the application is performed inside the TEE while all other operations (e.g., networking, UI) runs outside the TEE\footnote{Using TEEs for implementing trusted cryptographic functionalities for untrusted apps is mainstream -- e.g, KeyMaster\cite{keymaster} in ARM TrustZone.}.

This design has two advantages:   i) {\sys} can be seamlessly integrated with any E2EE messaging application without any change in messaging logic or specifications thus removing burden from the app developers, and ii) {\sys} enables security-conscious device vendors to provide privacy-enhanced communication capabilities by simply including the {\sys} library in their software distributions. Thus, 
we envision that {\sys} will be ubiquitously deployed \textit{even without additional support for hidden communication} (which we describe next). Consequently, having {\sys} on the device is not enough to raise suspicion unlike alternatives, e.g., deniable encryption, which have no other purpose for deployment other than hiding communication. 




In addition to trusted cryptography for E2EE messaging applications, {\sys} also enables "hidden messaging"; hidden messages sent using {\sys} can be later plausibly denied to the adversary. That is, a message transcript containing hidden messages is shown to be equivalent (to the adversary) to a message transcript that contains only messages sent through the messaging app which are revealed to the adversary. In this way, the adversary can verify that the observed message transcript indeed corresponds to inputs by the user into the messaging app (which may be adversary-controlled or monitored). 
This provides stronger deniability than ephemeral messages where no evidence can be provided to the adversary after the fact.


\paragraph{Hidden Messaging over Public Channel}
{\sys} establishes a "hidden communication channel" over the  public communication channel that is already established by the E2EE messaging application running in the potentially compromised environment outside the TEE. To achieve this, {\sys} injects (encrypted) hidden messages in the random coins (e.g., salts, IVs, etc.) inherently used in the public messaging cryptography. 
Specifically, encrypted hidden messages (indistinguishable from random) are used as random coins and then sent over the public channel masqueraded as randomly generated cryptographic metadata for the public messages. 
When coerced, users can hand over the encrypted public messages (along with the keys, and metadata) while still being able to plausibly deny the existence of hidden messages. 

Crucially, this approach does not impact the functionalities or alter the on-wire format of the messaging app. The E2EE app remains oblivious to the hidden message injection, and an adversary compromising the app by installing spyware or compromising the OS under which it runs has no visibility into the operations pertaining to hidden message injection. In fact, the way hidden messages are communicated over the public channel renders them "almost invisible" to the adversary. That is, the only indication of hidden messaging capabilities of the user's device is the installed {\sys} library. However, assuming that {\sys} is ubiquitously deployed, the presence of the {\sys} library is not sufficient to raise suspicion in itself, as all network traffic, execution transcripts, etc., look identical on all {\sys}-enabled devices \textit{regardless of whether it is being used for hidden communication}.

\paragraph{Integration with Existing E2EE Apps}
A {\sys} prototype implementation has been instantiated on {\tz} due to its widespread availability in mobile devices. {\sys} benchmarks show that overheads including those introduced due to context switches between the messaging application, the OS, and the TEE are in the order of milliseconds. In practice, \sys{}-introduced delays are completely overshadowed by user interaction. Further, injecting hidden messages does not incur communication overheads since the message transcripts do not increase in size over public messaging.

\section{Related Work}
\label{sec:related}
\begin{table*}[th!]
\centering
\small
\begin{tabular}{|llllll|}
\hline
\multicolumn{1}{|l|}{Scheme} & \multicolumn{1}{l|}{Property} & \multicolumn{1}{l|}{Dependency}  & \multicolumn{1}{l|}{\makecell{ Network \\ Adversary}} & \multicolumn{1}{l|}{\makecell{ Key \\ Disclosure}} & \multicolumn{1}{l|}{\makecell{ Device \\ Compr. }}  \\ \hline

\multicolumn{1}{|l|}{E2EE messaging} & \multicolumn{1}{l|}{Enc/dec on end-points, forward sec.} & \multicolumn{1}{l|}{-} & \multicolumn{1}{l|}{\cmark} & \multicolumn{1}{l|}{\xmark} &  \multicolumn{1}{l|}{\xmark} \\ \hline

\multicolumn{1}{|l|}{E2EE + ephemeral keys} & \multicolumn{1}{l|}{message/keys deleted periodically}  & \multicolumn{1}{l|}{-} & \multicolumn{1}{l|}{\cmark} & \multicolumn{1}{l|}{\cmark} & \multicolumn{1}{l|}{\xmark} \\ \hline

\multicolumn{1}{|l|}{OTR Messaging} & \multicolumn{1}{l|}{message content malleability} & \multicolumn{1}{l|}{IND-CPA secure malleable enc. } & \multicolumn{1}{l|}{\cmark} & \multicolumn{1}{l|}{\cmark} &  \multicolumn{1}{l|}{\xmark} \\ \hline

\multicolumn{1}{|l|}{Deniable Encryption} & \multicolumn{1}{l|}{ciphertext decrypts to any plaintext} & \multicolumn{1}{l|}{Indistinguishability Obfuscation} & \multicolumn{1}{l|}{\cmark} & \multicolumn{1}{l|}{\cmark} &  \multicolumn{1}{l|}{\xmark} \\ \hline

\multicolumn{1}{|l|}{Network Steganography} & \multicolumn{1}{l|}{obfuscation with cover traffic} & \multicolumn{1}{l|}{-} & \multicolumn{1}{l|}{\cmark} & \multicolumn{1}{l|}{\xmark} & \multicolumn{1}{l|}{\xmark} \\ \hline

\multicolumn{1}{|l|}{\textbf{{\sys}}} & \multicolumn{1}{l|}{hidden messaging over public channel} & \multicolumn{1}{l|}{TEE isolation} & \multicolumn{1}{l|}{\cmark} & \multicolumn{1}{l|}{\cmark} & \multicolumn{1}{l|}{\cmark} \\ \hline
\end{tabular}
\caption{A comparison of different flavors of secure/deniable messaging found in existing work. Property: flavor of deniability provided by the system, Dependency: dependency/assumptions on which the system is built, Net Adversary: secure against a network adversary monitoring communication transcripts, Key Disclosure: secure against key disclosure attacks, Device Comp.: secure against an adversary compromising the end devices. Except for {\sys}, none of the existing solutions provide any level of deniability on a compromised device. \label{tab:comparison}}
\end{table*}

\paragraph{End-to-End Encrypted Messaging}
 End-to-end encrypted messaging enables parties to communicate with each other via messages that are available in plaintext only on the end devices. When in transit, the messages are encrypted with keys that are available only to the parties. Most messaging apps today support E2EE messaging in addition to guarantees like forward secrecy (see Section \ref{sec:background}). While they are designed to provide orthogonal security guarantees, it is worth mentioning that end-to-end encrypted communication does not provide any deniability whatsoever once the encryption keys/metadata are revealed to the adversary.  
 
 \paragraph{Ephemeral Messages}
 Several messaging apps additionally support ephemeral messages where plaintext messages and keys/metadata are deleted periodically. While we can envision a solution for deniable messaging using ephemeral keys wherein keys are deleted after a message is sent/received, this does not provide strong plausible deniability guarantees: i) the messaging app may not be fully trusted/have backdoors, or may not duly delete the messages, ii) despite deletions it is possible for message artifacts and metadata to persist in the system e.g., in filesystem caches, and iii) if the device is already compromised, the adversary can observe messages in plaintext in I/O buffers, etc. This prompts the need for stronger mechanisms that are also resilient to device compromise.

\subsection{Plausible Deniability}

There is a long line of work building plausibly-deniable systems, mainly in the context of storage systems. This includes steganographic filesystems \cite{hanstegfs,pangstegfs,zhoustegfs,anderson98thesteganographic,petersGP15defy} and hidden volumes \cite{hive,pddm,datalair,chang2018mobiceal}. These solutions only work for data stored at rest and cannot be directly applied for deniable messaging. 
In the context of messaging applications, there are several flavors of plausible deniability proposed in existing work, and Table \ref{tab:comparison} compares the model in this work with existing solutions. We elaborate in the following.

\paragraph{Off-the-Record Messaging}
Off-the-record (OTR) messaging \cite{liu2013otr,borisov2004OTR} ensures equivocation for messages -- if Alice sends a message to Bob, she can later claim to a third-party (e.g., a judge) that Bob fabricated the message. To achieve this, OTR makes message ciphertexts malleable i.e., Bob could have generated a message from a potentially benign message that Alice sent. In this way, OTR protects against network adversaries with access to message transcripts, but if the end-point devices are compromised, the message contents and the ciphertexts are available to the adversary. 

\paragraph{Deniable Encryption}
 Deniable encryption \cite{canetti97deniable,canetti2020fullybideniable,sahaiSTOC14,oneill2011deniable} allows Alice and Bob to replay exactly a transcript of encrypted messages but to end up with a potentially different resulting plaintext than what was originally encrypted to present to an adversary. For example, even if the initial conversation between Alice and Bob was regarding a protest, the conversation transcript can be undetectably ``tweaked'' to show that the conversation was about attending a sports event. State-of-the-art techniques for deniable encryption rely on techniques such as indistinguishability obfuscation (IO) which are not practically realizable. In theory, it is unclear how IO can be realized securely on a compromised device, and even if this was possible, the adversary may obtain the messages in plaintext from other channels such as memory buffers, etc. %

\paragraph{Steganography}
Steganographic techniques, particularly designed to hide information in network traffic e.g., \cite{kaptchuk2021meteor} synthetically generate cover traffic to obfuscate the hidden information. For instance, Meteor uses a generative model to create cover text to hide information. While this suffices against a network man-in-the-middle adversary, on a compromised device, the cover text generation process as well as the encoding/decoding process is observable to the adversary. In addition, the techniques proposed in \cite{kaptchuk2021meteor} are not efficient enough to be applied in real-time to messaging apps. For instance, Meteor requires roughly 10 minutes to encode a 160-byte message, in addition to requiring up to 7KB of cover traffic. The throughput does not scale to messaging apps.

{\sys} uses random coins in symmetric key messaging applications to inject encryptions of hidden messages. In theory, this is similar to algorithm substitution attacks (ASA) where an adversary replaces an honest implementation of a cryptographic protocol with a subverted version \cite{bellare2014security,ateniese2015subversion,bellare2015mass}. Such substitutions of symmetric key cryptographic systems have been used before for the purposes of steganography \cite{berndt2017algorithm}. {\sys} can similarly substitute the cryptographic system used by the E2EE messaging application to support hidden messaging over a public communication channel. To realize this on a compromised device, where an adversary can inspect the cryptographic algorithms in use, the burden of implementation will be on the device vendors. However, since the cryptographic machinery used by E2EE messaging applications is often highly complex, designing such a system while retaining all privacy guarantees is non-trivial and will impose a significant implementation burden on the device vendors. As we will discuss later, {\sys} uses a simpler idea for injecting hidden messages that only needs to modify how random coins are selected for a symmetric key encryption scheme.



\subsection{TrustZone}
ARM processors support running applications in a TEE enforced by TrustZone. Inside this TEE vendors enable protecting security-sensitive applications (TAs) that run isolated from a potentially compromised REE (the Normal World). Previous work presents how TrustZone could be leveraged for protecting payment operations (TrustPay \cite{zheng2016trustpay}), providing one-time-passwords (TrustOPT \cite{sun2015trustotp}) and secure storage (DroidVault \cite{li2014droidvault}). Most of the proposed functionality has materialized in the form of TAs inside commercial TrustZone devices. Similarly, the \sys{} design aims to introduce a Secure World cryptographic library, which can also be used for hidden messaging.

Other work has focused on protecting users from Normal World adversaries. For example, TrustDump \cite{sun2014trustdump} enables collecting reliable memory snapshots, TruZ-View \cite{ying2019truz} provides trusted I/O paths to users, VeriUI \cite{liu2014veriui} provides attested login and SeCloak \cite{lentz2018secloak} provides reliable disabling of I/O devices from Secure World. Under TruZ-View, a user interface protects user input confidentiality from the untrusted OS. The VeriUI interface verifies user passwords inside Secure World, while the SeCloak interface enables users to enable and disable I/O devices from inside Secure World. \sys{}~provides a similar user interface, for entering hidden messages. Similar to SeCloak, the \sys{} interface takes over the display framebuffer to display hidden messages to the user. However, \sys{} goes further and also takes over the touch input in order to protect it against Normal World monitoring. The \sys{} provided interface enables users to inject hidden messages and read incoming ones, even under Normal World monitoring. 
\section {Background}
\label{sec:background}
\subsection{ARM TrustZone}
\label{background:TZ}

ARM Cortex processors enable building TEEs using the ARM TrustZone security extensions, or {\em TrustZone}. Under TrustZone, each physical processor core is split into two virtual CPUs. The processor either runs TEE software inside a {\em Secure World (SW)} or rich execution environment (REE) software inside the {\em Normal World}. The switching between the TEE and REE is controlled by a special {\em Non-Secure (NS) bit}. When the NS=0, then the core runs TEE code. Otherwise, REE software is executed. Physical memory regions and I/O peripherals are also tagged with an NS bit under TrustZone. Those tagged with NS=0 can only be accessed by Secure World, providing TEE exclusive control over the respective memory and I/O devices. The TEE can access all physical memory and dynamically allow or deny REE access to Secure World resources  (e.g., peripherals).

\paragraph{TrustZone communication}
The transition of control from Normal World to Secure World is known as a {\em world switch}. Both the REE and TEE can trigger world switches by issuing Secure Monitor Call (SMC) instructions. These instructions are handled by a {\em Secure Monitor}, which runs at ARM exception level EL3. 
Typically, regular applications running inside the REE and Trusted Applications (TAs) running under the TEE can also communicate with each other through OS-forwarded system calls as SMCs.

\subsection{Signal}
\label{wink:background:signal}

The Signal protocol  is designed with asynchronous messaging in mind -- messages can be sent even when the receiver is offline. This requires an intermediate server to facilitate key exchange with the help of information uploaded by each party during initialization. This key information for each user is stored as part of a {\em prekey bundle}, signed by the user, and is retrieved from the server when another user wants to establish a messaging session. 
 
The prekey bundle contains several types of keys which include: i) long-term {\bf identity keys}, ii) medium-term {\bf signed prekeys}, and iii) short-term {\bf ephemeral keys}.
Using these keys, a sender can establish a secure end-to-end encrypted messaging session with an offline receiver.

\section{Model}
\label{sec:model}

\paragraph{Plausibly-Deniable Messaging}
Informally, {\em plausibly-deniable messaging} (PDM) enables trusted peers to exchange messages but later plausibly deny the exchange to an adversary. Note that deniable encryption and off-the-record messaging, both of which provide some form of equivocation of the exchanged messages, also provide similar functionality. However, unlike these tools which mainly hide information in transit, PDM also hides all evidence (in network transcripts, systems artifacts, etc.) that a "hidden" message has been exchanged or a special system is being used with the sole purpose of hidden communication.

\paragraph{Deployment}
In a plausibly-deniable messaging scenario, a trusted  \textit{Sender} wants to send a message(s) to a trusted \textit{Receiver}. The sender and the receiver may later want to plausibly deny the exchange of certain messages to a powerful, coercive adversary (described next). For this purpose, both the sender and receiver use a plausibly-deniable messaging application on their respective devices.  Note that the application does not enable the sender and the receiver to  deny that (any) communication has taken place, but rather allows them to plausibly deny the contents of the conversation.

\paragraph{Plausible Deniability in Practice} 
For plausible deniability to be effective, we need to make a few general assumptions: 

\noindent\textit{\underline{Ubiquitous Deployment}:} 
First, if plausible deniability solutions remain a niche product, any user using such a system will appear suspect. Therefore, as with most works on plausible deniability, we will assume that {\sys} is ubiquitously deployed on mobile devices. 
In this regard, {\sys} arguably provides a strong use-case even when hidden messaging is not the primary purpose, i.e., 
as a sanitized environment for implementing cryptography of E2EE apps. This is unlike solutions like deniable encryption where the use-case beyond obfuscation of messages is unclear. 

\noindent\textit{\underline{Rational Adversary}:}
Second, as will describe later, {\sys} is deployed as a trusted service inside \tz. And since so far trusted services can only be installed by the device vendors, the inclusion of {\sys} in a system is completely in the hands of the vendor. This has an added benefit: if the vendor buys into the practicality of {\sys}, all devices by the vendor will have this trusted service.
Also, by design, only the vendor is able to extract information from the trusted service. This renders any amount of rubber-hose cryptography -- where the user is subject to confinement and torture -- useless since the user is unable to extract secret information from {\sys}. Thus, a rational adversary will need to approach the vendor through proper channels, e.g., with warrants, to access this information. In the recent past, this model has proven to be successful in protecting user privacy from overly intrusive government \cite{apple_fbi}.  Arguably, it is impossible to build a system resilient against adversaries that penalize the user irrationally regardless of evidence.

\noindent\textit{\underline{Deniability is Not Obfuscation}:}
Finally, it is worth noting that plausible deniability is \textit{not security by obfuscation}. The adversary may inspect the user device, analyze the binaries stored, etc., and identify that the software has features that enable deniable communication. The goal is to ensure that the adversary cannot detect (or guess with high probability) that the user ever uses these features. To achieve this, a plausible deniability system provides a plausible, alternative explanation for every action that is observed by the adversary in runtime.

\subsection{Threat Model}

\paragraph{Capabilities}
Since our adversary model is stronger than 
the adversaries considered in previous works (see Table \ref{tab:comparison}), we need to carefully define the powers of this adversary. In the following, we enlist these capabilities.

Specifically, we consider a coercive adversary who may:

\begin{itemize} [nosep,leftmargin=1.6em,labelwidth=*,align=left]
    \item Partially corrupt the sender's/receiver's devices and observe/record user inputs in plaintext.
    \item Analyze all software binaries, firmware, etc.,  on the device.  
    \item Observe, capture and store all (encrypted) communication for introspection. 
    \item Obtain access to both the sender and receiver devices (possibly at the same time) and coerce the users to hand over cryptographic keys.
\end{itemize}

Before discussing what the adversary can compromise on the user's device, it is worth clarifying a few points. First, we allow the adversary to examine the user device, i.e., analyze the binaries, firmware images, etc. In effect, the adversary knows that the {\sys} library is being used (at least to implement cryptographic operations for E2EE messaging apps), but as we will show, it does not know that the library is being used in some cases for  additional hidden messaging. Thus, the goal is not to hide the existence of the {\sys} library in the system but rather to make the modes of operation indistinguishable to the adversary. 

Second, the adversary may compel the user to hand over keys. As will describe later, {\sys} has a set of keys that are accessible through user-chosen passwords. Some of these keys are \textit{public} and are handed over to the adversary and some are \textit{hidden} which the user plausibly denies using. Due to the {\sys} design, the hidden keys are never provided to the user in the plain. The user is only able to use the hidden key through the {\sys} library based on a password input. When coerced, the user denies having a password that enables the hidden key(s) for hidden messaging operations.



\paragraph{System Setting \& Device Compromise}
{\sys} runs in a trusted execution environment (TEE) (namely TrustZone). Therefore, standard TEE-based security assumptions hold:

\begin{enumerate}[nosep,leftmargin=1.6em,labelwidth=*,align=left]
	\item TEE isolation can not be compromised via software or device hardware vulnerabilities. 
	\item The software running inside the TEE is trusted. 
	\item REE cannot overwrite the TEE set device configuration. 
	\item {The device vendor is trusted and there are no TEE backdoors.} 
\end{enumerate}

\paragraph{Attack Vectors}
Based on the different privilege levels, there are four broad categories of adversarial attack vectors.

\begin{enumerate}[nosep,leftmargin=1.6em,labelwidth=*,align=left]

    \item {\em Compelled disclosure:} The adversary coerces the user to hand over message transcripts, which includes the plaintext messages and the corresponding ciphertexts sent and received, and the related 
    cryptographic metadata (e.g., keys, random coins, etc.). 

    \item {\em Compromising REE Apps (Normal World App Compromise)}: The adversary installs keyloggers \cite{keyloggers}, spyware \cite{memory_spyware_without_root} or even have backdoors into the messaging app. Using a compromised app, the adversary may try to violate message confidentiality, capture screenshots or key types and monitor the user's actions, with the goal of detecting hidden payloads in the exchanged messages.

    \item {\em Compromising the REE Kernel (Normal World Root Access):} The adversary leverages the compromised kernel into monitoring from Normal World operations performed by \sys{}. Monitoring can include tracking TEE entries and exits, timing \sys{} calls, or detecting I/O resource usage. Only a (hijacked) REE kernel would be capable of monitoring and reporting such fine-grained TEE operations.

    \item {\em Compromising the TEE (Secure World Compromise):} The adversary leverages TEE software or hardware vulnerabilities to escalate their privileges inside the TEE. Once inside the TEE, the adversaries have direct access to the  plausibly-deniable messaging application. Further, TEE compromise might provide adversaries with complete control over the device.  
\end{enumerate}

This paper mainly focuses on defending against attacks in categories 1 and 2, which hold even under the assumptions and are easily deployed i.e. without requiring advanced exploits or alerting the user. Attacks in category 4 are out-of-scope as it immediately invalidates all assumptions and effectively give the adversary complete control over the device. Denial of service attacks e.g., intentionally blocking incoming/outgoing messages, etc. are specifically considered out-of-scope of the threat model. Note that all plausible deniability systems are vulnerable to such DoS attacks. Designing DoS-resilient plausible deniability schemes is an open problem.

\paragraph{Side Channels}
As with any shared-hardware system {\sys} may be subject to side-channels that undermine certain security guarantees. We have identified two such side-channels that break plausible deniability provided by {\sys}: i) a timing-based channel that measures the time of process execution inside the TEE by tracking entry and exit inside the TEE, and ii) detecting Secure World screen-utilization by monitoring voltage, display frequency changes or other indicators of change in displayed images.
We have addressed these side-channels in the design provided in Section \ref{sec:winkHeavy}.
While we acknowledge that there may be other side-channels that are detrimental to the security of {\sys}, exploring and mitigating them is the subject of ongoing work, and orthogonal research \cite{liu2020cost,prime_count}.


\paragraph{Cryptography}
To make {\sys} practical, it is desirable to only employ standard (and efficient) cryptographic primitives. Specifically, {\sys} only requires the following: i) the existence of a secure public-key infrastructure (PKI) with trusted  certificate authorities, ii) the existence of efficient mechanisms to jointly compute shared secrets between two trusted parties, and iii) the existence of cryptographically-secure one-way functions e.g., cryptographic hashes, KDFs, etc.

\section{The \sys{} Design}
\label{sec:winkLite}

 This section details the {\sys} design. As discussed before, {\sys} requires a TEE-enabled device. The current version is instantiated under ARM TrustZone, the most commonly available TEE environment for commercial mobile devices. TrustZone provides all the TEE security guarantees listed in Section \ref{sec:model}. Dozens of sensitive applications (TAs) are already protected by TrustZone on ARM commercial mobile devices (e.g., KeyMaster, SamsungPay, WideVineDRM, etc.) by running inside the Secure World.

An E2EE messaging app has several key components including networking, user I/O, and cryptographic operations. The most obvious (yet impractical) design of a "hardware-assisted secure world messaging app" for TrustZone-enabled devices runs all components of the app under Secure World protection. While hidden messaging is straightforward in this case, since without compromising the Secure World the adversary has no visibility into the messaging process, there are at least three obvious problems with this design. First, adding an entire messaging app codebase to Secure World exponentially amplifies the TCB. For instance, the Signal and Telegram codebases have over 257 KLOC and 791 KLOC respectively, while open-source Secure World OSes have codebases that typically do not exceed 220 KLOC (e.g., OP-TEE kernel, Nividia TEE and LinaroTEE all contain less than 210 KLOC). Second,  networking components can be exploited to gain control over the Secure World TAs or the OS. 
Finally, since TAs can only be installed by the device vendor, realizing this design would require collaboration between the app developers and the vendors.


Therefore, in {\sys}, only the most critical components of the messaging process are executed inside the TEE. The challenge is to identify these components based on the adversarial capabilities. In this Section, we present a design where the adversary under consideration may compromise the messaging app running in the Normal World but is not capable of compromising the Normal World kernel through code injection, ROP, etc. For this threat model, only executing the cryptographic operations in the Secure World suffice to realize hidden messaging. 
The following Section deals with an adversary capable of compromising the Normal World kernel.

 \begin{figure*}[t!]
 	\centering
 	\includegraphics[scale=0.75]{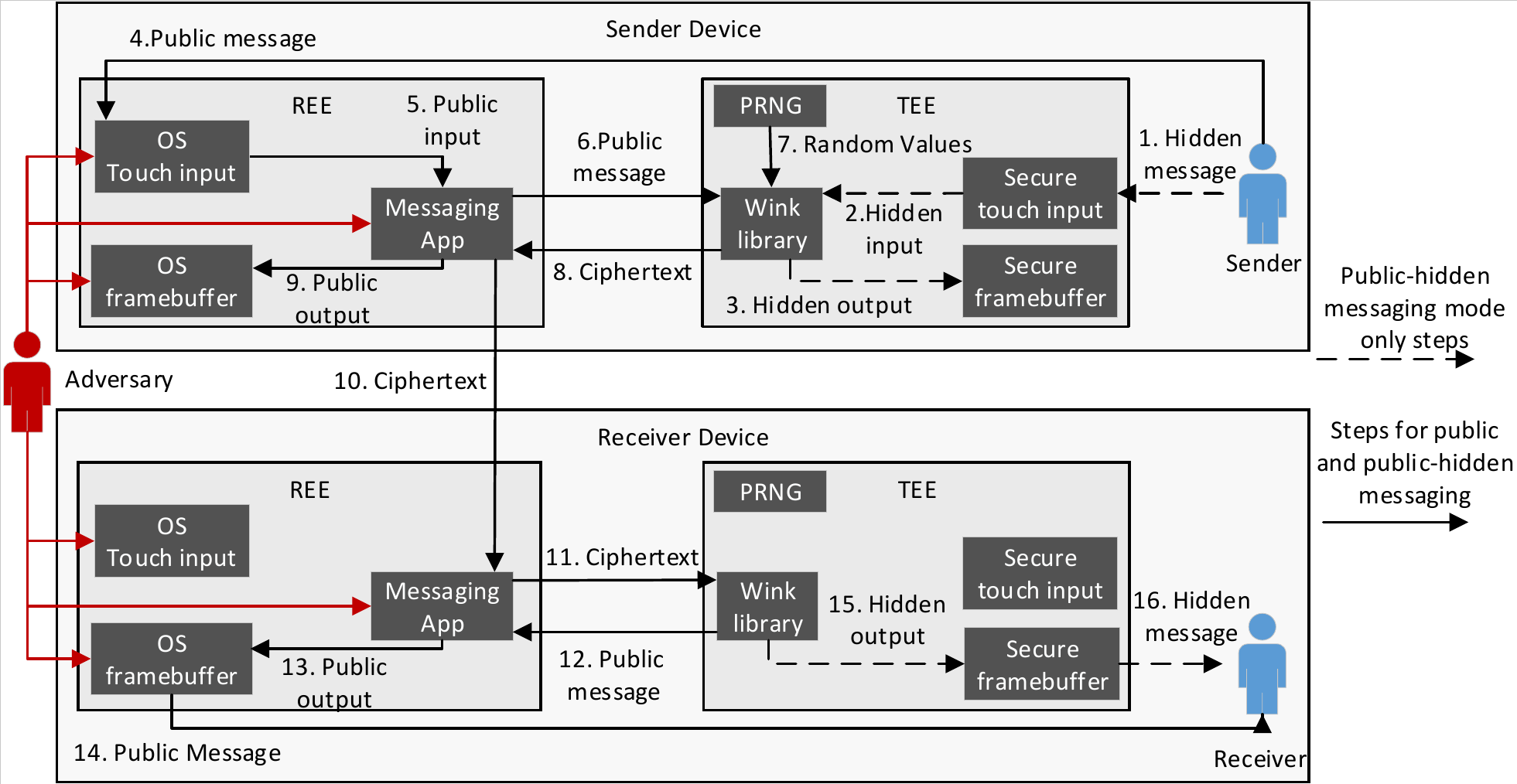}
 	\vspace{1pt}
 	\caption{The \sys{}~framework. {\sys} consists of a TEE hosted library that provides cryptographic implementations for E2EE messaging apps and secure I/O for providing users with a hidden messaging interface. The REE remains oblivious to any injected hidden messages.}
 	\label{fig:wink_framework}
 \end{figure*}


\subsection{The \sys{} Secure World Application}
\label{operating-modes}

Under TrustZone, {\sys} runs as a trusted application (TA) inside the Secure World and implements the cryptographic primitives used by the Normal World messaging app to provide a secure authenticated channel for public messaging. This is established using PKI keys whereby after an initial certificate exchange, a symmetric session key is jointly computed e.g., using the Diffie-Hellman Key Exchange protocol. The resulting {\em public session key} encrypts public messages in that session. Importantly, all cryptographic keys are protected by the TEE and the encryption/decryption of public messages is performed within the TEE, isolated from the untrusted software running in the REE.
Compared to running E2EE messaging apps entirely in the REE, this provides stronger security guarantees for the cryptographic information, making {\sys}-integrated apps more resilient to unauthorized key disclosures, etc. Figure \ref{fig:wink_framework} illustrates the design. 

In commercial TrustZone devices, TAs cannot be installed directly by the users. Instead, only vendor-signed software can execute as TAs inside Secure World. It is standard for the vendor to provide security-critical services such as \sys{} for users as TAs running inside Secure World.  {\em Importantly, since users cannot install \sys{}, the presence of the corresponding library on a user device does not indicate in any way to the adversary that it is installed for the purpose of hidden messaging.} Further, most TrustZone device vendors already provide crytographic libraries similar to the one \sys{}~proposes as TAs (e.g., KeyMaster). The key difference between these libraries and \sys{} lies in the ability to provide hidden messaging.

\paragraph{Modes of operation}
{\sys} has two modes of operation: a {\em public-only} messaging mode and a {\em public-hidden} messaging mode. The user should typically operate the device in the public-only mode and use the public-hidden mode when safe. A user password input decides the mode of operation. {\sys} requires three passwords: i) a \textit{public} password corresponding to the public-only mode, ii) a \textit{hidden} password corresponding to the public-hidden mode, and iii) a \textit{disclosure} password for verification of public communication (described later). The ceremony for setting up the passwords is sketched in Section \ref{setup-ceremony}. When the Normal World messaging app is started, it requires the user to provide a password. The password input and verification are performed directly within the Secure World. The password determines the mode of operation.

\begin{itemize}[nosep,leftmargin=1.6em,labelwidth=*,align=left]
\item {\bf Public-only}. In this mode, {\sys} only performs 
cryptographic operations for the messaging app in the Normal World. 
{\sys} operates in this mode on the public password input. {\sys} stores public message-specific metadata (which may also contain hidden data) until the messages are deleted by the messaging app. This metadata is stored in order to allow verification of public communication later under coercion (described later).

\item {\bf Public-hidden}. In this mode, {\sys} may also inject hidden messages into the public communication channel. This mode is enabled only upon the provision of a hidden password. The existence of the hidden password is denied under coercion. In this mode, users may also input and view hidden messages through a Secure World-provided user interface (described next). Encrypted hidden messages are injected into the public messaging channel while those already received are displayed on the interface.  \sys{} protects hidden message confidentiality and only shows them to users upon hidden password input. Thus, it is crucial for the user to not reveal the existence of the hidden password to adversaries either accidentally or under coercion. A leaked hidden password can lead to revealing the public-hidden mode usage as well as the hidden messages exchanged.


\end{itemize}

\subsection{Secure I/O for Hidden Messaging}
\label{trustzone:hidden-io}

{\sys} protects hidden messages both in transit and on the end devices themselves. For this, no evidence of hidden I/O is leaked Normal World software, including the messaging app itself and the Normal World OS. To accomplish this, the \sys~cryptographic library sets up a hidden messaging interface by directly communicating with the touchscreen through Secure World-protected I/O channels. 

In order to securely display information to the user, \sys~takes control over the {\em touchscreen framebuffer driver} when the user holds the power button for three seconds. The power button interrupt is configured for Secure World usage such that \sys{} first receives all power button presses. This setup ensures that users can trigger the public-hidden mode interface without Normal World knowledge. Of course, regular button presses are forwarded to the Normal World for maintaining standard power button functionality.  
Once the library running in public-hidden mode receives this interrupt, first \sys~temporarily prevents the Normal World from reading or writing framebuffer data. Then, the library saves the last framebuffer state and displays the hidden messaging interface. The hidden messaging interface is drawn on the framebuffer from inside the Secure World, displaying a keyboard and the exchanged hidden messages. Additionally, a user-specified watermark is also shown on the interface to prevent interface spoofing attacks. This watermark is maintained inside Secure World and its setup is detailed in \ref{setup-ceremony}. 

For user input, \sys~takes control over the touch input device and monitors for user touches from inside the Secure World. The user input monitoring starts once the hidden messaging interface is displayed. The monitoring stops once the user exists the interface by triggering the \sys{} hardware interrupt. The user-provided input is then communicated as one or more hidden message chunks.
Both user input and output are hidden from the Normal World, as the input buffers are always cleared and the framebuffer restored prior to returning executing in Normal World.

\subsection{\sys{} Setup Ceremony}
\label{setup-ceremony}

\begin{figure}[h]
	\centering
	\includegraphics[scale=0.63]{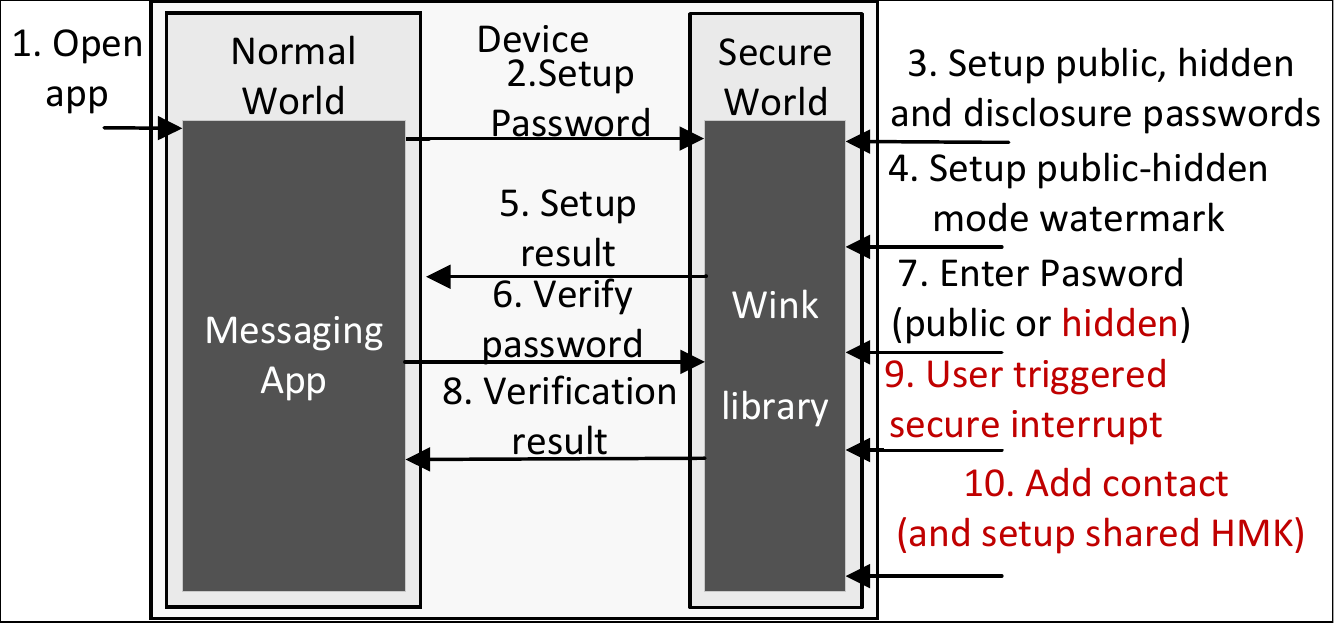}
	\vspace{1pt}
	\caption{\sys{} setup ceremony. For a {\sys}-integrated messaging app, the Secure World hosted {\sys} library manages the password setup and verification. Entering the public-hidden mode password enables setting up hidden contacts and hidden message exchanges using symmetric encryption keys dubbed Hidden Master Keys (HMKs).}
\label{fig:wink_setup}
\end{figure}

Figure \ref{fig:wink_setup} depicts the ceremony of setting up the {\sys} library. Initially, the user asks the messaging app to set up her password. This request is forwarded to the \sys{} library, which will provide the user with an interface for entering the three passwords (public, hidden, and disclosure passwords). Once the user has entered her passwords, \sys{} will also ask the user to enter a text only she knows. This text is drawn on the hidden messaging interface as a watermark in order to prevent spoofing attacks from the Normal World. Once all required data is entered \sys{} returns execution to the messaging app.

Subsequently, the password hashes and the watermark never leave the Secure World. Instead, \sys{} indicates a successful verification of the messaging on the provision of either of the public or the hidden password. In both cases of operation, the information relayed to the messaging app is exactly the same, ensuring that the app is oblivious to the actual mode of operation. Only If the hidden password is provided hidden messaging functionalities are enabled in {\sys}.

\subsection{Interfacing with Messaging Applications}



\begin{figure*}[ht]
	\centering
	\includegraphics[scale=0.78]{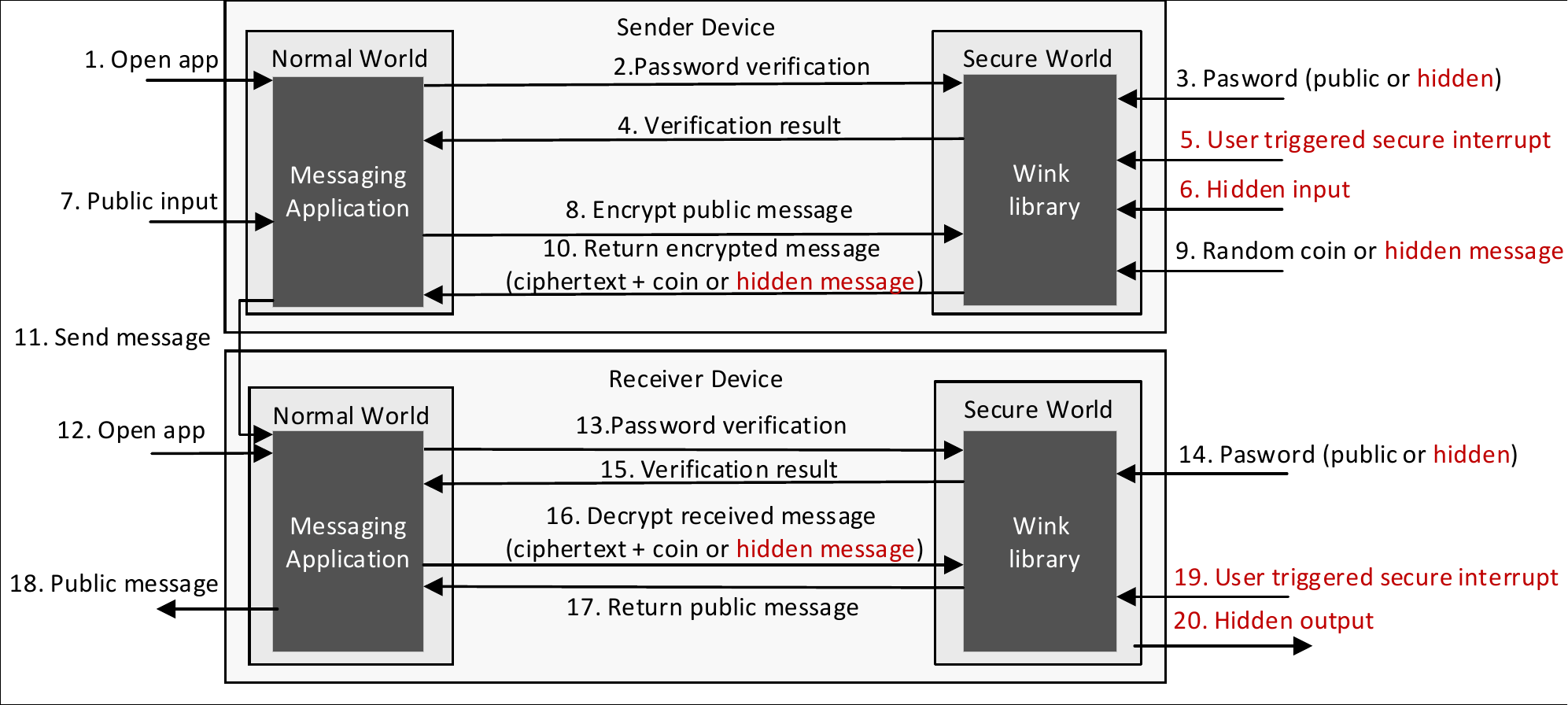}
	\vspace{1pt}
	\caption{Message exchange under \sys{}. Based on the user password, {\sys} runs either in the public-only or the public-hidden mode. In public-hidden mode, the user can enter and read hidden messages, which are encrypted and exchanged between devices masqueraded as the random coins of public messages. }
\label{fig:wink_comm}
\end{figure*}

Figure \ref{fig:wink_comm} provides a high-level overview of the interactions between the Normal World messaging app and the {\sys} library (e.g., setting up passwords, adding contacts, and exchanging messages). Specifically, the Normal World applications (messaging applications) can only access the library through a set of APIs provided for handling cryptographic operations and storing sensitive information (e.g., keys, and passwords). These APIs are exposed through a series of SMCs, which can be accessed by the messaging applications through OS-provided system calls. In order to use the provided APIs, the application just has to call the corresponding system call. All system calls and parameters are visible to the OS, which forwards them to the library inside Secure World. Thus, it is crucial that no evidence of hidden messaging is leaked by the parameters passed or the calls to the cryptographic library themselves.

The Secure World hides the inner workings of \sys~cryptographic operations from both Normal World applications and the OS. Once the Secure World receives an SMC call, it obtains complete control over the device and can perform the hidden operations required by \sys{}. These operations include (i) reading and writing hidden user input and (ii) injecting hidden messages into the public communication channel.  The public and hidden message exchanges under \sys{} are illustrated in Figure \ref{fig:wink_comm}. The black steps are executed in both public-only and public-hidden mode, while the red steps are only executed in public-hidden mode. Importantly, the generated random coin depicted (step 9) is overwritten by the hidden message ciphertext chunk in public-hidden mode. 

 Note that sending each hidden message chunk requires an accompanying public message. Further, the hidden message can only be reconstructed and read by the receiver when sufficient hidden message chunks are received for its reconstruction. Thus, to deliver a hidden message, the sender has to construct and send as many public messages as there are hidden message chunks. The number of chunks created by {\sys} for each hidden message depends on the bandwidth available, determined by the message format of the public messaging app, and is discussed in more detail in Sections \ref{eval:message-bandwidth} and \ref{implementation:telegram}. However, the sender can always construct and send additional public messages when they are required to finish the delivery of a hidden message. \sys{} helps the sender plan his public messages by noting how many are required to finish sending the current pool of hidden messages.



\subsection{Key \& Contact Management}
\label{trustzone:hmk}
The {\sys} Secure World TA securely stores all cryptographic information required for messaging (both public and hidden) encrypted under keys derived from a device-specific master key. Both the master and derived keys never leave Secure World, ensuring that only \sys{} knows the derived keys. Similarly, \sys{} also stores encrypted contact information and Hidden Master Keys (HMKs) for hidden messaging, which are saved encrypted on physical persistence storage. Ideally, the dedicated TEE persistent storage would be used for saving these details. However, not all TrustZone-enabled commercial devices provide such storage. Thus, this information is stored in the form of an encrypted blob at a fixed Normal World location, a standard mechanism for securely storing TA data. 

A pre-determined amount of persistent storage is reserved for saving HMK and contacts when {\sys} is installed. The entire blob is read into memory when the device boots up and there is a call to the {\sys} library, and thereafter re-encrypted and stored during a graceful shutdown. {\em This ceremony is performed regardless of the presence of any hidden keys, metadata, etc.} Naturally, due to the fixed capacity, the number of contacts that can be added for hidden messaging is limited, but since the only information {\sys} requires per contact is their corresponding HMKs, a small amount of storage may suffice. For instance, with 64-bit user identifiers and 1024-bit HMKs, 1MB of reserved storage will allow more than 3000 contacts. We also note that storage-level plausible-deniability is a well-studied problem, even in the context of mobile devices \cite{chen2020infuse,jia2017deftl,petersGP15defy}. These solutions can be employed here (in future versions) for more efficient designs.

\paragraph{Hidden Keys \& Contacts}
To facilitate hidden messaging, for each contact, users are required to exchange a shared secret {\em once} over an out-of-band communication channel. This may be realized either through physical exchange when the users meet or through cryptographic protocols like deniable authenticated key exchange \cite{raimondo2006DAKE,Unger2018ImprovedSD}. In either case, the shared secret never leaves the Secure World. In case the users exchange the secret in person, it is input directly into the Secure World (using secure I/O) and is made available only on the provision of a correct hidden password. For this, when the messaging app adds a new contact, the \sys{} library generates the secret and a corresponding QR code. The QR code is then presented as a secure output from the Secure World. The counterpart {\sys} TA (being added as a contact), scans this QR code as input into the Secure World. For this, Secure World takes control of the camera temporarily and loads the QR image. Finally, the secret is extracted from the QR image and maintained inside Secure World, associated with the corresponding contact. This exchange process is similar to the physical verification process in several existing messaging apps. For example, under Signal key bundles are verified also by scanning QR codes. Under \sys{}'s much more powerful threat model, this optional feature becomes mandatory for hidden messaging due to the lack of a trusted communication path prior to the exchange of the HMK.

The shared secret is used to derive a hidden master key (HMK). All hidden messages 
are encrypted under the HMK. Even a fully compromised Normal World can not access the \sys{} HMKs maintained inside Secure World. Without having access to the HMK, the adversary cannot breach the confidentiality of the hidden messages {\em even if the public keys are revealed}. In the current design, each pair of communicating apps have a unique HMK, which is securely stored and encrypted with a password-derived key by the Secure World. All HMK(s) are made available to the \sys{} library only upon the provision of the correct hidden password. As a result, providing the hidden password makes all HMKs stored on the device available. In addition, {\sys} has a static HMK per user which is used for the hidden communication channel.
Future designs will include more fine-grained access controls, and --
at the cost of additional bandwidth -- also possibly additional security features such as forward secrecy i.e., the HMK refreshed through key ratcheting and KDFs, etc.

\paragraph{Controlled Disclosure of Public Keys}
 Protecting public message keys in the TEE makes the public messaging opaque to the adversary. However, public message keys and public metadata have to be disclosed in accordance with the standard plausible deniability model where a user is asked to hand over a key by a coercive adversary (see Section \ref{sec:model}). In several nation-states, handing over encryption keys is in fact law-mandated \cite{ukencryptionlaws}.  Since the public message key(s) is stored in the Secure World, we need a mechanism to reveal this on-demand to the user, without making the keys accessible to compromised apps or OS.  Without Secure World compromise, the only way to retrieve such information is through \sys~provided APIs. {\sys} supports controlled disclosure using a disclosure password. On provision, {\sys} reveals the encryption keys and metadata for past public messages (up to the point stored by the messaging app) using secure output. Once the disclosure password has been used, {\sys} allows a password reset.

Only users in possession of a {\em disclosure password} (i.e. those also in possession of both hidden and public passwords) can use the provided API for disclosing public messaging details. The disclosure password enables \sys{} to provide a transparent public messaging mode, where users can disclose all public messaging details to adversaries upon coercion. However, \sys~ensures that no users, even under coercion, can ever disclose the hidden messaging details. Importantly, the keys are revealed only using Secure World output and are not exposed to any Normal World software.

\subsection{Security Analysis}
\label{sec:winkLite:secAnalysis}

The security arguments underlying {\sys} fall into three categories. First, it needs to be argued that for any generic E2EE messaging application that is compatible with {\sys}, integration does not come with lower security assurances. In other words, a {\sys}-enabled messaging application retains the same message privacy guarantees that the messaging application has without {\sys}. Second, it needs to be shown that injecting hidden messages into public communication does not impact the privacy guarantees of public communication, while also ensuring that the message injection remains undetectable in the message transcripts. Informally, message transcripts correspond to information exchanged on-wire which includes the message ciphertexts, random coins, etc. Finally, {\sys} should also ensure that any normal world monitoring (short of kernel compromise) cannot detect hidden message injection based on execution patterns. Execution transcripts correspond to on-device operations when {\sys} operates in a particular mode of operation which includes data inputs, context switches, the timing of operations,  etc.

In the following, we show that the message transcripts of a {\sys}-enabled E2EE application is indistinguishable from the message transcripts of the same application without {\sys} to a computationally-bound adversary observing the network transcripts. In other words, {\sys} does not distinguishably alter the message transcripts of the E2EE application, and thereby does not compromise the privacy guarantees of the vanilla application. We also show that for a more powerful device-resident adversary in the Normal World that is incapable of privilege escalation, the message transcripts of {\sys} operating in the public-only mode and public-hidden mode are indistinguishable. Finally, we show that the Normal World adversary cannot distinguish modes of operation based on the execution patterns.




\begin{restatable}{theorem}{nonCompromise}
    \label{thm:winkLite:non_compromise}

    Given a generic E2EE messaging application \genericEToEE, compatible with {\sys}, and a {\sys}-enabled version of \genericEToEE, denoted \winkEToEE, operating in either public-only or public-hidden mode, the {\bf message transcripts} of \genericEToEE and \winkEToEE are indistinguishable to a PPT adversary if IND-CPA secures encryption producing ciphertexts indistinguishable from random exists. 

\end{restatable}

\begin{restatable}{theorem}{messageTranscript}
\label{thm:winkLite:messageTranscript}
Normal world adversaries (incapable of privilege escalation) and on-wire adversaries cannot distinguish between {\bf message transcripts} pertaining to public-only communication and transcripts containing public + hidden communication for a {\sys}-enabled E2EE messaging app if there is an IND-CPA secure encryption producing ciphertexts indistinguishable from random exists.
\end{restatable}

\begin{restatable}{theorem}{executionTranscript}
\label{thm:winkLite:executionTranscript}
Normal world adversaries (incapable of privilege escalation) cannot distinguish between {\bf execution transcripts} pertaining to either public-only or public-hidden modes of operation.
\end{restatable}

Proofs are in \ref{sec:proof-1}, \ref{sec:proof-2} and \ref{sec:proof-3}. 


\section{{\sys} Under REE Kernel Compromise}
\label{sec:winkHeavy}

In this section, we first describe the impact of REE kernel compromise on the \sys{} design presented in Section \ref{sec:winkLite}. Then, we describe a solution unifying the public and hidden I/O channels through the Secure World preventing confidentiality breaches due to REE kernel compromise.  
Finally, we analyze the impact of this solution on the TCB size, ease of integration with messaging apps, and overall plausible deniability.

\paragraph{Impact of Kernel Compromise}
Events in the past \cite{CVE-2014-4322, CVE-2014-4323,CVE-2015-6640} have shown that REE kernel vulnerabilities may enable sophisticated privilege escalation from compromised REE applications and provide adversaries with control over the REE kernel. Taking over the kernel will not only provide direct access to all public messages but also enable tracking and timing of TEE operations, as mentioned in Section~\ref{sec:model}. Specifically while \sys{} ensures that hidden message confidentiality remains intact despite REE kernel compromise, the interrupts used to initiate hidden message inputs can be monitored by the compromised REE kernel. {\sys}-introduced interrupts due to user I/O may be arbitrarily long (although typically in the order of seconds). Thus, it is straightforward for a compromised kernel to time \sys{} Secure World execution and determine when it handles interrupts specific to \sys{} hidden I/O. Padding interrupt handling to the order of user I/O would render such monitoring useless. However, it is infeasible due to the impact on system performance.

\paragraph{Unifying Public and Hidden Interfaces} Without moving the entire messaging app logic into the Secure World, one solution to tackle this problem is to obfuscate the hidden message I/O time with the public message I/O time. Specifically, in addition to the porting public messaging cryptography, \sys{} simply needs to provide a UI for public messaging directly using the Secure World I/O, alongside the hidden interface detailed in Section \ref{sec:winkLite}. In this way, the time required for public-only I/O can also be used for additional hidden I/O. Using the same interface for both public and hidden I/O is also more practical and user-friendly since it enables the user to view and write public and hidden messages concurrently.  Hardware interrupts are no longer required to initiate hidden inputs rendering any Normal World monitoring useless. 

Handling public I/O through the TEE also has an added advantage: \textit{message confidentiality is now ensured against a compromised REE kernel}. More specifically, since the REE now only sees the messages encrypted and essentially acts like a "man-in-the-middle", confidentiality is guaranteed when using an appropriate IND-CPA secure encryption scheme. However, it is worth noting that this does not subsume the need for hidden messaging since ensuring confidentiality is not enough for plausible deniability. Coercive adversaries will still demand cryptographic metadata pertaining to the messages exchanged, and to maintain plausible deniability, the user now hands over the plaintext public messages, and the cryptographic metadata used to encrypt these messages using the controlled disclosure feature detailed in Section \ref{sec:winkLite}.

\paragraph{Cost of Integration}
The added security of porting the public I/O to the Secure World comes at the cost of increased integration complexity and a larger TCB. In particular, each messaging app has a unique user interface for messaging, which can vary in complexity and code size depending on the features supported by the app client (e.g., pictures/movies, emojis, voice messages, etc). 
Integrating and maintaining UI for each app into \sys{} requires a major engineering effort and increases the Secure World TCB, perhaps unreasonably in some cases. In contrast, {\sys} only requires a minimalist implementation of an I/O interface for hidden messaging since the low bandwidth hidden communication channel offered by most messaging apps can only support text messages. Clearly, this interface is not enough for everyday public messaging needs with modern messaging clients. 

Ideally, if multiple messaging apps were to subscribe and use a generic Secure World UI provided by the vendor, then the increase in TCB may be reasonably controlled in addition to making app integration and maintenance straightforward. We believe that security-conscious messengers have enough incentives to adopt this model 
in collaboration with device vendors given the added protection it offers for public messages. With messaging apps being frequently targeted 
to breach confidentiality via system compromise~\cite{kismet,pegasus,vultur}, a more secure E2EE app needs to be resilient to these attacks.

\paragraph{Security Analysis}
Unifying the I/O interfaces for public and hidden messages does not alter the message transcripts generated in the public-only and public-hidden modes of operation. Without obtaining the cryptographic metadata for the hidden messages protected in the Secure World, the adversary cannot distinguish (with more than negligible advantage) message transcripts for public-only communication and public + hidden communication (Theorem \ref{thm:winkHeavy}). Compromising the REE does not enable the adversary to compromise the Secure World in our threat model. 
We show that with the proposed changes to the {\sys} design, a compromised REE kernel also cannot distinguish the mode of operation based on the execution transcripts.

\begin{restatable}{theorem}{winkHeavy} 
\label{thm:winkHeavy}
A compromised REE kernel cannot distinguish between {\sys}'s public-only and public-hidden modes of operation based on their corresponding {\bf execution transcripts} if IND-CPA secure encryption produces ciphertexts indistinguishable from random exists. 
\end{restatable}

The proof sketch is in \ref{sec:proof-4}.

\paragraph{Side-Channels}
As with any shared-hardware system, the threat of side-channels is applicable to {\sys}. However, unifying the public and hidden message interface eliminates all hidden interface-specific resource usages. Thus, the REE adversary is prevented from using any potential side-channel based on these resource usages. 

As described before, timing entry and exits into the Secure World does not provide any additional information. Since the interface for inputting/outputting hidden messages and public messages is the same, the resource utilization due to screen usage is also indistinguishable between the cases when the screen is used only for public messaging vs. public + hidden messaging. Thus, monitoring resources such as power consumption, refresh rate changes, etc. due to the UI does not reveal any information to the adversary. That is, any variations observed in these parameters are as likely to be the result of only public message input/output as due to both public and hidden message input/output. The intermixing of public and hidden messaging operations inside the TEE.

Obviously, if additional hidden messages are processed inside the Secure World, additional computation is required. However, very few additional cycles are used for injecting the hidden payload in the random coins (as demonstrated in Section \ref{wink:evaluation}). Power consumption does not vary out of noise boundaries. Thus, while side-channels based on computation or other resources might be discovered in the future, it is not immediately clear if these side-channels can in fact be exploited to break plausible deniability (see Section~\ref{wink:discussion} for further discussion). Discovering and mitigating such side-channels is indeed very important, and each side-channel mandates its own evaluation in terms of effectiveness and evaluation. However, this is beyond the scope of this paper.
\section{Implementation}

Only device vendors have access inside the TEEs of TrustZone commercial mobile devices and this access is strongly guarded. Instead, we have implemented the first {\sys} prototype on a i.MX6 Nitrogen6X Max \cite{imx6board} development board, featuring a mainstream ARM Cortex-A9 CPU and 4GB of DDR3 memory. More importantly, the board provides complete access to both Normal World and Secure World. 

The \sys~prototype uses U-boot \cite{denk2013u} to load up a minimal operating system, OP-TEE \cite{optee} inside the Secure World and a vanilla Android 7.0 Nougat inside the Normal World. When the device powers on, U-boot first loads into memory the code of both OP-TEE and Android. Then, U-boot passes execution to OP-TEE, which starts executing with complete control over the device. OP-TEE sets up the Normal and Secure World regions, assigns interrupts for each, and prepares handlers for inter-world communication. Then, OP-TEE sets up drivers for Secure I/O and prepares a cryptographic library for handling Normal World requests. Once the Secure World is set up, OP-TEE passes execution to the boot code of the Android kernel (Linux version 4.1) inside Normal World.


The {\sys} trusted computing base (TCB) consists only of code executing inside Secure World. In the prototype, Secure World only contains a stripped-down version of OP-TEE (12769 LOC), drivers for the user I/O (4423 LOC), and the {\sys} cryptographic library (3009 LOC). In total, the Secure World only contains 20201 lines of code.

\subsection{Secure World cryptographic library}

As described in Section \ref{sec:winkLite}, only a Secure World TA is required for injecting hidden messages into an authenticated secure public messaging channel. For simplicity, the prototype TA runs as part of the OP-TEE which forwards all SMCs incoming from Normal World to the {\sys} cryptographic library API. 
%
%
The library can operate in either public-only or public-hidden mode (Section \ref{operating-modes}). The library initially starts in public mode, where it can (i) encrypt/decrypt messages, and (ii) save cryptographic keys and passwords. On password storage requests from the messaging app, the user can either enter only the public and disclosure passwords or also a hidden one through Secure I/O. On password verification, the library switches into public-hidden mode only when provided the hidden password. 
As described in Section \ref{operating-modes}, in public-hidden mode the library can also show the hidden messaging interface and injects hidden messages into the messaging app's public messaging channel. {\em Crucially, \sys{} ensures that the duration of processing encryption/decryption requests in public-hidden mode is indistinguishable from when they are processed in public-only mode.}



When injecting a hidden message, the library notifies the user how many public messages are required to successfully send the entire hidden content. In the current implementation, for each public message, only a 15-byte hidden message chunk can be embedded for Telegram (Section \ref{implementation:telegram}) and a 16-byte chunk for Signal (Section \ref{signal-integration}) . The library keeps injecting the hidden message chunks until they are all exhausted. When no hidden messages are provided by the user, the library operates similarly to the public-only mode. On the receiver side, the library extracts the hidden messages and displays them when operating in the public-hidden mode.

\subsection{Secure I/O}
\label{implementation:io}

Not unlike cell phone chipsets, the i.MX6 provides a set of registers inside the Central Security Unit (CSU). These registers control the accessibility of peripherals, including I/O devices. \sys~ takes advantage of this in order to on-demand enable and disable Normal World access to the touchscreen input (I2C touch interface) and output (display framebuffer).

\paragraph{Output} 
{\sys} displays an interface for hidden messaging when \sys{}-specific hardware interrupt is triggered by the user. For our implementation, we have chosen to set up a Secure World GPIO key driver that intercepts the power button interrupts when \sys{} operates in public-hidden mode and shows the hidden messaging interface if this button is pressed for 3 or more seconds. Because the Secure World only intercepts the power button hardware interrupts, Normal World software can not trick \sys{} into showing the respective interface. Further, users can only trigger the hidden messaging interface provided they know the \sys{} hidden password. Temporary altering the power button functionality also prevents the user from accidentally shutting down her phone, while the \sys{} library operating in public-hidden mode has not sent all hidden messages or shown all received ones. 

The hidden messaging interface always displays the exchanged hidden messages and a static keyboard, which can be used by the user for input. A similar interface is displayed for entering and verifying passwords.
A framebuffer driver inside Secure World is used to draw the two interfaces. Using its control over CSU configuration, the driver takes over the touchscreen display and its framebuffer while the library is running. If the power button is pressed while \sys{} operates in public-hidden mode, the framebuffer driver first saves the current Normal World framebuffer image. Then, the hidden messaging interface is drawn on the framebuffer and displayed to the user. The Normal World image is restored in the framebuffer prior to returning execution to the Normal World. This process hides the framebuffer driver operations from Normal World adversaries.

Normal World adversaries could try to trick the users into entering hidden messages or passwords into a spoofed hidden messaging interface. To prevent such attacks, the framebuffer driver always displays the user-entered text watermark (at setup time) on the framebuffer. This text is provided using Secure I/O and never revealed to the Normal World, informing the user when the authentic hidden messaging interface is shown by the Secure World.

\paragraph{Input}
The Secure World provides hidden input to the cryptographic library by directly monitoring the I2C interface used by the touchscreen to send touch input data. While the hidden messaging interface is displayed, Secure World takes control of the I2C interface and prevents Normal World adversaries from monitoring the interface for hidden messages.

On touch events, the I2C interface provides the Secure World driver with a buffer that indicates touch screen presses and the coordinates of each touch. On touch presses, the driver converts the received data into a 2D point.
This point is then mapped into a static keyboard layout displayed on the touchscreen Secure World framebuffer. On each touch event, the driver provides the entered keystrokes to the {\sys} library.

\section{Evaluation}
\label{wink:evaluation}

\paragraph{Integration with Messaging Apps}
\label{wink:implementation:integration}
%
%
As proofs of concept, we have integrated the {\sys} library with Telegram Messenger (see Appendix for detailed results) and Signal Private Messenger. In addition, we have investigated several other messaging apps including Briar\footnote{\url{https://briarproject.org}}. For Telegram we have identified a 15-byte salt being exchanged under the MTProto 2.0 protocol described in the Appendix. On the other hand, the Signal message format provides two opportunities for hidden message injection: a 32-byte random ECDH public key exchanged with every message, and a 16-byte IV for the Sealed Sender encryption (an envelope) for exchanged messages. Finally, the Briar encrypted message is exchanged using metadata that includes a 32-byte random salt and 24-byte IV. 

In our evaluation, we mainly present empirical results pertaining to the solution presented in Section \ref{sec:winkLite} which protects against REE adversaries incapable of compromising the REE kernel. The only change required for making the solution resilient against REE kernel adversaries is performing public I/O through the Secure World interface which does not introduce additional timing 
overheads. 

\setlength{\tabcolsep}{1em}
\begin{table}[t]
\caption{Bandwidth available for hidden messaging with different E2EE messaging apps. With Telegram, for a $n$ byte hidden messages, $n/15 + 2$ public messages are required as cover traffic. Using the Signal 16 byte IV a similar bandwidth is available. Using both the 32-byte salt and 16-byte IV provides a significantly larger bandwidth for hidden message injection under Briar. }
\small
\vspace{4pt}
\label{apps:bw}
\centering
\begin{tabular}{@{}llll@{}}
\toprule
\makecell{Application}                       & \makecell{Random Coins } & \makecell{No. of Public \\ Messages} \\ \midrule
Telegram \cite{telegram} & 15 byte salt               & $\ceil{\frac{n}{15}} + 2$               \\
Signal \cite{signal}                        & 16 byte IV                & $\ceil{\frac{n}{16}} + 2$               \\
Briar \cite{briar}                         & 32 byte salt, 16 bytes IV & $\ceil{\frac{n}{48}} + 2$               \\ \bottomrule
\end{tabular}
\end{table}



\begin{table*}[th!]
\centering

\begin{tabular}{@{}|c|c||c|c|@{}}
\toprule
              & Signal & {\sys}-Integrated Signal & Overhead\\ \midrule
Metadata Encryption time (ms)    & 0.17 ± 0.003  & 0.42 ± 0.002 & + 0.246 ms  \\ \midrule
Metadata Decryption time (ms)    & 0.19 ± 0.004  & 0.43 ± 0.003 & + 0.240 ms  \\ \bottomrule
\end{tabular}

\caption{{Encryption and decryption time for metadata (33 bytes) in Signal ± {\sys}. Overheads are under 0.25 msec per message. This is virtually unnoticeable and dominated by other per-message operations which take {\bf thousands} of times longer (e.g., typing, network transfer, etc.). \label{tab:signal-eval}}}
\end{table*}

\paragraph{Bandwidth for Hidden Messaging}
\label{eval:message-bandwidth}
The length of the random coins determines the available bandwidth 
available for hidden message injection in each messaging app. Table~\ref{apps:bw} enlists the available bandwidths for the surveyed apps. 
For Telegram we have built a prototype (evaluated in Section \ref{implementation:telegram}) that enables sending a $n$-byte hidden message using $\ceil{\frac{n}{15}} + 2$ public messages, where the 2 extra messages are for the metadata. Similarly under Signal sending a $n$-byte hidden message requires $\ceil{\frac{n}{16}} + 2$ public messages (details in Section \ref{signal-integration}). Briar provides both a 32-byte salt and 16-byte IV which combined enables {\sys} to inject additional hidden message bytes per public message.

\subsection{Integration with Signal}
\label{signal-integration}
As a proof-of-concept, we have integrated the Android Signal Private Messenger with {\sys} (see Section~\ref{sec:background}). The instrumented Signal enables users to exchange hidden messages under the Wink protocol while retaining all existing functionalities for public messaging. For Signal, \textbf{60 LOC} have been changed for hooking in the Wink system calls, mostly consisting of Java Native Interface code.

\paragraph{Switching Operating Modes}
In terms of user authentication, Signal does not maintain passwords or require users to explicitly log in. Instead, it mainly relies on 2FA (through SMS sent to provided phone numbers) and OS-level user authentication. Thus, we had to find an appropriate method of hooking into the password verification required under \sys{}. Instead of introducing mandatory authentication every time Signal is opened, we have opted to hook into the Signal PIN functionality. The PIN is a (numeric/alphanumeric) code used to support features like non-phone number-based identifiers. 

We instrument the Signal PIN verification and move it into Secure World under \sys{}. This enables users to provide passwords (Signal PIN) to the \sys{}~library. In turn, the library provides verification results back to Signal. Similarly, the PIN input is managed through the \sys{}~library allowing the setup of the three \sys{}~ passwords (public, hidden and disclosure).

Once the \sys{} hidden password is set up, the user can use Signal's PIN verification to enable the \sys{} public-hidden mode execution. As previously described in Section \ref{implementation:telegram}, once the hidden password is entered, the Secure World intercepts all power button presses and shows the \sys{} hidden messaging interface, and entered hidden messages are exchanged while appearing as random Signal IVs.


\paragraph{Hooking Hidden Messaging} 
%
For Signal, we inject the hidden messages into the IV used for encrypting message metadata under the sealed sender \cite{signal-sealed-sender} functionality. In short, a sealed sender adds another layer of encryption (an envelope) to each exchanged message. The metadata IV for encrypted envelopes can be replaced with any random string, thus making it suitable for hidden message injection. We leverage this opportunity to integrate {\sys} by porting the envelope (metadata) encryption inside Secure World, similar to the Telegram integration described in the process described in Section \ref{telegram-hidden-message} for Telegram. Thus, for each Signal exchanged message, the metadata is encrypted and exchanged using either a {\sys} generated random IV or a hidden message encrypted with CTR mode AES under a 16-byte randomly generated HMK.

To instrument the Signal PIN verification and to introduce hidden message exchange we only had to replace a few lines of code with system calls. However, only message metadata encryption is ported inside Secure World for this proof-of-concept. For a real-world deployment, further engineering effort would be required to also move the remaining Signal cryptographic functions used for message content encryption inside Secure World for protection.

\label{eval:processing}

\paragraph{Signal Metadata Encryption Overheads}
Under the sealed sender functionality, Signal encrypts and decrypts the metadata of each exchanged message using AES in CTR mode with no padding. For our {\sys} proof-of-concept we have ported these two operations inside the Secure World library. To estimate the overheads for integrating {\sys} with Signal we compared the time required for encrypting/decrypting the fixed length metadata of exchanged messages between the vanilla and {\sys}-integrated Signal.

To evaluate the metadata encryption/decryption times we instrumented Signal to automatically encrypt/decrypt the metadata of each message sent/received 1000 times. The average AES encryption and decryption are measured, collecting the average computation time and standard deviation. 
 Table \ref{tab:signal-eval} presents the average metadata encryption and decryption time. AES encryption of metadata introduces an overhead of under 0.246 milliseconds for either encryption or decryption (of metadata -- always 33 bytes), with a 95\% confidence interval. The overhead includes the additional data copy operations from the Normal World to the Secure World. Importantly, most other operations take thousands of times longer (typing, message transfers, etc). As a result {\sys} overheads are virtually unnoticeable.




\section{Discussion}
\label{wink:discussion}

In this section, we further discuss threat model assumptions and point out potential {\sys} design drawbacks. 

\paragraph{TEE Compromise}
Naturally, the TEEs themselves are far from invulnerable. Importantly, however, the small, comparatively carefully designed TEE TCB results immediately in proportionally fewer vulnerabilities, oftentimes by 2-3 orders of magnitudes!
%
\cite{Cerdeira2020SoKUT} shows that between 2013 and 2018 only a handful of CVEs have been reported for TEEs, in comparison to the 1647 CVEs reported for Linux, a rich OS. Additionally, in Trustonic, one of the most largely-deployed TEEs, only 5 vulnerabilities have been reported in this timeframe, {\em 330 times} fewer than for Linux (1647). Moreover, most reported TEE vulnerabilities are located within TAs, which are isolated by the TEE OS from \sys{} and do not directly impact hidden messaging security. 

Overall, the fact is that the significantly smaller TCB/attack surface raises the bar significantly. However, hardening TEEs and specifically Trustzone against all possible attack vectors is orthogonal to the scope of this work. 


\paragraph{Side Channels}
As with any shared-hardware system {\sys} may be subject to side-channels that undermine its security guarantees. As described before, the solution in Section \ref{sec:winkHeavy} mitigates several side-channels that are specifically detrimental to {\sys} security guarantees. However, we acknowledge that there may be side-channels that are not covered or mitigated under the current {\sys} design. There is a large body of work attempting to make progress in mitigating these side-channels for general \tz-protected trusted services, e.g., \cite{sectee,cagalyst,hybcache,liu2020cost}. Deploying this proposed solution will eliminate corresponding side-channels detrimental to {\sys} security. In addition, identifying other side-channels that may undermine {\sys} is the subject of future research.

\paragraph{Password-based attacks}
As with any password-based system, users are assumed to use strong passwords, which can not be guessed in a reasonable amount of time. To prevent password guessing, \sys{} imposes strict length and content requirements and locks access permanently after a reasonable amount of attempts.
%

%

%

\paragraph{Spoofing \sys{} UI} 
Adversaries could try to show users a spoofed version of the \sys{} interface (e.g., from a compromised application), to trick them into leaking their passwords (hidden and public) or hidden messages. Such attacks can be mitigated in various ways, including by the user-selected watermark maintained only inside Secure World and shown on the \sys{} drawn interface.

\paragraph{Data leakage through Wink} 
\sys{} is preferably installed by a vendor as firmware in the Secure World and provides a secure communication channel that is invisible to the Normal World. It only becomes visible to users when they utilize it through the \sys{} hidden-messaging UI. Importantly, the vendor is trusted by the user. Secure World is assumed to be benign and not compromised.

However, it is important to understand what that means. To that end, let us consider a malicious vendor or a Secure World OS that compromises \sys{} operations. In addition to obtaining all confidential information exchanged as hidden messages, they can also introduce their own collected data (e.g., collected user confidential information) into hidden messages and exfiltrate it without the user or Normal World software ever knowing. Thus, a compromised \sys{} can become a powerful tool for adversaries if they require a mechanism for leaking data from Secure World without leaving any evidence. Note, however, that such adversaries can introduce their encrypted data within any other exchanged random coins within Normal and Secure World software and achieve a similar result. \sys{} does not introduce such data leakage as a new attack vector.

\section{Conclusion}
This work presented {\sys}, a plausibly-deniable messaging framework enabling users to reclaim the ability to communicate securely even in the presence of powerful surveillance or coercive adversaries. It works by surreptitiously injecting hidden messages in cryptographic randomness inherent in end-to-end encrypted messaging. Users can plausibly deny the exchange of hidden messages, and also any evidence of using the messaging software itself. {\sys} can be efficiently integrated with a number of existing E2EE applications including Telegram and Signal with minimal external instrumentation, and crucially without needing to change existing standard on-wire message formatting.


\section*{Acknowledgments}

This work was supported in part through NSF award 2052951. We thank the anonymous Usenix Security Symposium reviewers for their 
excellent suggestions and feedback. The views and conclusions in this document are those of the authors and should not be interpreted as representing the official policies, either expressed or implied, of the National Science Foundation.


\begin{small}
\bibliographystyle{splncs04}
\bibliography{bib/crypto.bib,bib/deniability.bib,bib/misc.bib,bib/trustzone.bib,bib/full.bib}
\end{small}
\appendix
\renewcommand{\thesection}{\appendixname~\Alph{section}}

\begin{figure*}[th!]
\centering
\label{fig:encryption_timing}
 \subfigure[Encryption Time]{
\label{fig:wink_encryption}
\begin{tikzpicture}[scale=0.8]
\pgfplotsset{%
    width=.540\textwidth,
    height=0.36\textwidth
    }
\begin{axis}[
  ybar,
  ymin = 0,
  bar width=7pt,
  xlabel=Message size,
  xtick = {1,2,3,4,5,6,7},
  xticklabels = {1,4,16,64,256,1024,4096},
  ylabel=Time (ms),  
  y label style={at={(0.08,0.5)}},
  legend pos=north west,
        enlarge x limits={abs=0.5},
        	cycle list = {black, black!50!white, black!90,black!40,black!10,}
]

\addplot+ [fill, text=black, error bars/.cd,
y dir=both,y explicit
        ] coordinates {
            (1,1.054309) +- (0,0.0442)
            (2,1.051362) +- (0,0.0639)
            (3,1.046617) +- (0,0.0534)
            (4,1.058508) +- (0, 0.034)
            (5,1.070808) +- (0,0.0433)
            (6,1.152979) +- (0,0.0318)
            (7,1.436233) +- (0,0.0394)
        };

 \addplot+ [fill, text=black!50!white, error bars/.cd,
y dir=both,y explicit
        ] coordinates {
            (1,1.995651) +- (0,0.0848)
            (2,2.001823) +- (0,0.084)
            (3,2.007989) +- (0,0.087)
            (4,2.007114) +- (0,0.085)
            (5,2.054234) +- (0,0.0817)
            (6,2.229802) +- (0,0.0858)
            (7,2.868889) +- (0,0.0546)
        };
\legend{Telegram, {\sys}-Integrated Telegram}
\end{axis}
\end{tikzpicture}
}
\subfigure[Decryption Time]{
\label{fig:wink_decryption}
\begin{tikzpicture}[scale=0.8]
\pgfplotsset{%
    width=.540\textwidth,
    height=0.36\textwidth
    }
\begin{axis}[
  ybar,
  ymin = 0,
  bar width=7pt,
  xlabel=Message size,
  xtick = {1,2,3,4,5,6,7},
  xticklabels = {1,4,16,64,256,1024,4096},
  ylabel= Time (ms),
  y label style={at={(0.08,0.5)}},
  legend pos=north west,
        enlarge x limits={abs=0.5},
        cycle list = {black, black!50!white, black!90,black!40,black!10,}
]

\addplot+ [fill,text=black, error bars/.cd,
y dir=both,y explicit
        ] coordinates {
            (1,0.984477) +- (0,0.0391)
            (2,0.982414) +- (0,0.028)
            (3,0.981982) +- (0,0.0275)
            (4,1.003036) +- (0,0.0483)
            (5,1.025316) +- (0,0.0413)
            (6,1.127767) +- (0,0.0302)
            (7,1.538629) +- (0,0.0302)
        };

 \addplot+ [fill, text=black!50!white, error bars/.cd,
y dir=both,y explicit
        ] coordinates {
            (1,1.948663) +- (0,0.0391)
            (2,1.952700) +- (0,0.0427)
            (3,1.950715) +- (0,0.0387)
            (4,1.965746) +- (0,0.0455)
            (5,2.005862) +- (0,0.0291)
            (6,2.167213) +- (0,0.0306)
            (7,2.837355) +- (0,0.0357)
        };
        
\legend{Telegram, {\sys}-Integrated Telegram}
\end{axis}
\end{tikzpicture}
	}	
\caption{Comparison of encryption, decryption times for {\sys}-integrated Telegram and vanilla Telegram. The error bars illustrate the margins of error for a 95\% confidence interval. Performing Secure World cryptographic operations with the {\sys} library incurs a 2x overhead.}
\end{figure*}
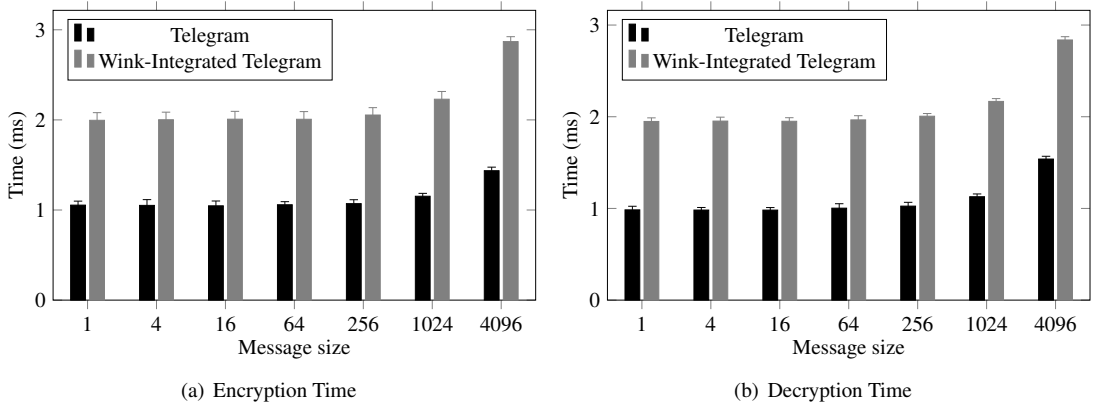

\section{Integration with Telegram}
\label{implementation:telegram}
We have integrated Telegram with {\sys}. Telegram is a popular open-source E2EE messaging app based on the MTProto 2.0 protocol. Compared to 
its more complex counterparts e.g., Signal, one particular advantage of the Telegram protocol is that the cryptographic operations are relatively simple to implement. 
The following describes the salient features, more details can be found in \cite{telegram}. 

\paragraph{Message Format}
MTProto 2.0 derives a per-message 16 byte {\em message key} by hashing a shared {\em authentication key} and 
an augmented message payload, consisting of the message text itself, a 64-bit session identifier, a 64-bit salt (15 bytes in the actual implementation) and the required length of the padding. The message 
key is used to derive AES keys and IVs for the payload encryption using a key-derivation function (KDF). The encrypted payload, message key, and authentication key are sent over the wire. The salt provides the necessary randomness to the message keys and the payload. As will be discussed later, this is also used to inject hidden messages using {\sys}.

The instrumented version retains all existing functionalities for public messaging. Only \textbf{71 LOC} have been changed for hooking in the \sys{} system calls, mostly consisting of added Java Native Interface code and replacing the existing encryption/decryption logic. Importantly, since the Telegram message format is not altered, the instrumented version can still communicate with a vanilla Telegram version for public messaging, while relying on the \sys{} library for cryptographic operations.




\paragraph{Instrumenting Encrypted Messaging Logic}
The \sys{} library stores authentication keys (shared key used in MTProto 2.0 protocol) inside Secure World and performs public message encryption and decryption. To this end, we instrumented the Telegram "secret chat" encrypted messaging logic using the MTProto 2.0 protocol. This logic mainly consists of deriving message keys and IVs through a series of SHA256 (or SHA1 for MTProto 1.0 protocol) computations, followed by an AES IGE~\cite{aes_ige} encryption or decryption. Under \sys{}, this logic is replicated inside the Secure World library.


\paragraph{Switching Operating Modes}
For user authentication, Telegram allows users to set up a passcode (or PIN). Under \sys{}, Secure World library SMC calls replace Telegram passcode entry and verification logic. As a result, when passcode protection is enabled, the user is prompted by the \sys{} to enter all three passwords (public, hidden, and disclosure), through a Secure I/O interface.  
These passwords are never revealed to the Normal World and the \sys{} library only relays to Telegram whether password verification succeeded or failed. 



On providing the hidden password, the Secure World intercepts all power button presses and shows the \sys{} hidden messaging interface every time the power button is pressed for longer than 3 seconds. The interrupt-based provided interface enables users to manage hidden contacts and read/write hidden messages at any point during the \sys{} public-hidden mode execution. The interception of the power button presses ends when \sys{} resumes operating in public mode (e.g., by entering the public password). Hidden messages are then exchanged as random Telegram message salts.


\paragraph{Telegram Hidden Message Injection}
\label{telegram-hidden-message}
Telegram's MTProto 2.0 protocol includes an 8-byte salt in the encrypted message sent over the wire. In the open-source code, we found that the actual salt size used is 15 bytes (perhaps to make Telegram more resilient to brute-force attacks). {\sys} uses this 15-byte salt for hidden message injection. 

Hidden messages are encrypted with AES in CTR mode under a 16-byte randomly generated hidden master key. A hidden message, if longer than 15 bytes, is encrypted as a whole and the ciphertext is broken into 15-byte chunks. Each chunk is subsequently used as a random salt for a public message and sent over the wire as part of the encrypted payload. 

In addition, each hidden message is preceded by a 16-byte random IV and a 8-byte message length indicator, which is broken in two 15-byte encrypted messages. The remaining 6 bytes out of the combined 30 bytes is used to add redundancy e.g., all 0s allowing the decryption logic to distinguish between actual messages and random information. 

While this simple scheme judiciously utilizes the available bandwidth (by reducing the amount of metadata) it has one drawback: if the chunks for a long-hidden message are received out-of-order then the decryption logic on the receiver side may not be able to reorder the encrypted ciphertext chunks. We leverage the fact that each encrypted payload in Telegram has a {\em sequence number} which allows the receiver to correctly re-order out-of-order public messages. 

Since the hidden message injection in {\sys} is synchronized with the public messaging i.e., each hidden chunk is injected with one public message, the ordering of the public messages also establishes the hidden message chunk ordering. Thus, {\sys} can leverage the public message sequence numbers to correctly reorder the hidden message chunks for decryption.

\paragraph{Telegram Message Processing Overheads}
\label{eval:telegram-processing}
To estimate the performance overheads of integration, we compared the time required for encrypting/decrypting messages in  {\sys}-integrated Telegram with the time required in the vanilla version. Note that the only difference between the two cases is that the Secure World-hosted {\sys} library implements the crypto for the {\sys}-integrated version while the vanilla version implements its own cryptography in the Normal World. 

The experiment is set up by instrumenting Telegram to automatically generate and encrypt/decrypt messages of sizes between 1 and 4096 bytes (max size supported by Telegram). For decryption, a remote Telegram instance encrypts and provides messages between 1 and 4096 bytes. For each generated message the experiment is repeated 1000 times, measuring the encryption process -- SHA256 computations for key and IV generation plus AES IGE encryption -- duration. Then, the average execution time and standard deviation are computed. A similar process is carried out for timing the decryption of message ciphertexts. 



Figure \ref{fig:wink_encryption} and Figure \ref{fig:wink_decryption} illustrate the average time required for encrypting and decrypting messages under both vanilla and \sys{}-integrated Telegram respectively. 
The observed results indicate that using the {\sys} library for secure cryptographic operations only makes the encryption, and decryption slower by a factor of 2x. This is expected given the context switches and the memcopies required for utilizing the {\sys} cryptographic operations. 
For a messaging application, the overhead is almost negligible since it only slightly increases overall message processing times (around 1 ms). The bulk of the time is spent on user inputs, which are usually in the order of a few seconds. The results also indicate that the time required for encryption and decryption is not significantly affected by the message size. {\em Note that the almost minimal overheads justify using {\sys} only for more secure public messaging and strengthen plausible deniability arguments.}

\section{Security Proofs for Section \ref{sec:winkLite:secAnalysis}}
%



\subsection{Message Privacy in E2EE Apps}
Most E2EE messaging apps aim to provide message confidentiality, integrity, and forward security. Typically, applications such as Signal and Telegram use authenticated encryption schemes that provide both authenticity (integrity) of messages and confidentiality.

\paragraph{Confidentiality}
Authenticated encryption schemes that are block-cipher based, such as AES in GCM mode of operation, are \textit{nonce-based symmetric encryption} schemes \cite{rogaway2004nonce}. These schemes aim to produce message ciphertexts that are indistinguishable from random. Specifically, they satisfy $IND\$-CPA$ security (note the difference from standard $IND-CPA$ security) against a PPT adversary \adv -- i.e., given access to i) an encryption oracle $\encrypt(\kappa,.,.)$ which returns ciphertexts encrypted under the encryption scheme \genericEncryptionSchemeDef for arbitrary messages selected by \adv, and ii) an oracle $\$(.,.)$ which returns random bits of the same length as a ciphertext under the encryption scheme, \adv can distinguish between outputs of these oracles with only negligible advantage. Mathematically, 

\[
\prob{\kappa \getsr \keyBundle: \adv^{\encrypt(\kappa,.,.)} \implies 1} - \prob{\adv^{\$(.,.)} \implies 1} \leq negl(\secParam)
\]

where $\keyBundle$ is the key space from which keys are drawn for \genericEncryptionScheme.

To achieve $IND\$-CPA$ security, block ciphers require a random nonce per message (which is captured by \genericRandCoin in Definition \ref{defn:etoee}). There are multiple ways in which the random nonce is deployed. In applications where AES is used in CBC mode or GCM mode, \genericRandCoin corresponds to the IV/nonce required for encryption. In other applications, \genericRandCoin may be used as a random seed into a key derivation function (KDF) to provide a random key for encrypting each message (e.g., Signal, see below). Finally, \genericRandCoin can be used as a salt that is prepended/appended to a message before encryption (e.g., Telegram, see below).

\paragraph{Integrity}
Message integrity in an authenticated encryption scheme is guaranteed by attaching a \textit{tag} that is usually a message authentication code (MAC). The MAC is either computed on the plaintext data which is then encrypted alongside the plaintext ("MAC-then-Encrypt"), or on the ciphertext and the tag is attached to the payload ("Encrypt-then-MAC"). Importantly {\bf the tag generation process is not determined in any way by the choice of the random nonce}.

\paragraph{Forward Security}
Informally, forward security in messaging applications ensures that messages that are sent/received in an "epoch" are not accessible to an adversary after the epoch has passed. This guarantees that an adversary that corrupts a device at a given time is not able to break the confidentiality of previously exchanged messages. In particular, messaging applications like Signal operate in sessions (or epochs) where a \textit{chain key} is used to derive a set of random symmetric keys that are used for encrypting/decrypting messages in that session (e.g., using the chain key as a seed to a random oracle). Once that session is over, the chain key is deleted, and a new chain key is established using a key ratcheting protocol. Since the chain keys for previous sessions are no longer accessible to an adversary that compromises a device in the current session, the keys used for encrypting/decrypting the messages are also not accessible to the adversary. 


\paragraph{Properties Preserved by {\sys}}
{\sys} preserves message confidentiality, integrity, and forward security when integrated with an E2EE app. Showing that a {\sys}-enabled E2EE application retains integrity and forward security guarantees is straightforward: {\sys} only alters the way a random nonce is selected for the authenticated encryption scheme used in the E2EE app, which {\bf impacts neither message integrity (tag generation is independent of the random nonce) nor forward security (which entirely depends on the key derivation mechanism)}. Thus, as long as the E2EE app implements {\em message integrity} using an authenticated encryption scheme, and provides {\em forward security} through its key derivation process
a {\sys}-enabled version of the app retains these two properties. 

Showing that {\em message confidentiality} is also retained requires further analysis, described in the following.

\subsection{Definitions}
To show that {\sys} retains also confidentiality, we first need to define essential end-to-end encrypted messaging functionality and associated security guarantees. 

Modern messaging applications are highly complex. Multiple components are working in tandem including networking, UIs, cryptographic logic, etc. Here we will mainly discuss cryptographic functionality. 


A formal, all-encompassing definition of all the cryptographic functions required/supported in a messaging application is tricky. A recent attempt by Canetti \etal \cite{canetti2022universally} makes significant progress in this direction. Specifically, their work proposes modular definitions for an E2EE application in the universal composability (UC) framework. At a more technical level, the definitions are geared towards the Signal (or Signal-like) protocol that provides authenticated encryption, and forward security through key ratcheting. 

The paper defines an ideal functionality for secure messaging that implements two functions $\mathsf{SendMessage}$ and $\mathsf{ReceiveMessage}$, which simply produce encrypted (and resp. decrypted) messages using an authenticated encryption scheme. These functions also interact with key derivation functions that produce per-message keys while ensuring forward security. Additionally, there are functions for setting up long-term keys on the message servers. 

\paragraph{Focus is required for clarity and correctness}
While the definition of Canetti \etal is comprehensive 
its use-case is somewhat orthogonal and not particularly well suited to the current task at hand.
Specifically, by construction

\begin{itemize}[nosep,leftmargin=1.6em,labelwidth=*,align=left]
    \item {\sys} does not alter the setup of long-term keys on the message server for a {\sys}-enabled E2EE application
    \item {\sys} does not alter the authenticated encryption scheme used for producing encrypted messages in a {\sys}-enabled E2EE application.
    \item {\sys} does not alter the long-term and short-term key derivation mechanisms in a {\sys}-enabled application.
\end{itemize}

In fact, all the critical E2EE messaging application protocols remain untouched after integration with {\sys}. Thus, functions that are not impacted (e.g., setting up long-term keys, key derivation, etc.) can be defined abstractly. This allows us to more effectively focus on the key functions for 
sending and receiving messages, \sendMessage and \recvMessage which take as input a message, obtain a message key by making black-box use of the key derivation function \keyDerive, and produce an output using the authenticated encryption scheme \genericEncryptionScheme. 

There are several advantages to this focused approach. Firstly, the security implications of {\sys} can be analyzed through straightforward reductions to standard cryptographic assumptions in a game-based framework 
without proving security in the UC framework. Secondly, this allows us to capture the impacts of integrating {\sys} with any generic E2EE app, without relying on the fact that the messaging application is compliant with the UC definitions proposed by Canetti \etal \cite{canetti2022universally} for Signal-like protocols. Indeed, {\sys} can be integrated with any messaging app that provides random coins for injecting messages, and not just applications that provide Signal-like functionalities. 
Finally, this also allows us to "zoom in" and analyze the consequences of the (somewhat minor) changes that {\sys} makes to the overall cryptographic workflow of a complex messaging app, without re-defining all the functionalities of the app.






\begin{defn}
\label{defn:etoee}
An end-to-end encrypted messaging application client, denoted by $\genericEToEE$, makes black box use of an $IND\$-CPA$ secure authenticated encryption scheme, \genericEncryptionSchemeDef, and implements a set of algorithms $(\registration, \keyDerive, \sendMessage, \recvMessage)$

\begin{itemize}
    \item $\registration(\secParam, \clientID)$ takes as input a security parameter \secParam, and a client identifier \clientID, and outputs a set of "long-term" keys 
    $\keyBundle \defeq \{\kappa_1, \dots, \kappa_n\}$ where for $i \in [1, n]$, $\kappa_i \in \{0,1\}^\secParam$.

    \item $\keyDerive(\secParam, \keyBundle)$ takes as input a security parameter \secParam, and the set of long-term keys, and returns a key $\kappa \in \{0,1\}^{\secParam}$.

    \item $\sendMessage(\genericMsg, \clientID, \kappa, \genericRandCoin)$ takes as input a message \genericMsg, the client identifier of the message recipient, \clientID, a key $\kappa$ derived using \keyDerive from the key space of \genericEncryptionScheme, and a random string $\genericRandCoin \in \{0,1\}^{\secParam}$. 
    
    The function returns $\encryptedMsg = \encrypt(\kappa, \genericMsg, \genericRandCoin)$ where \genericRandCoin is a random coin (e.g., an initialization vector for a block cipher) used in the \encrypt algorithm. 

    \item $\recvMessage(\encryptedMsg, \kappa, \genericRandCoin)$ takes as input an encrypted message \encryptedMsg, a key, $\kappa$, and a random coin $\genericRandCoin$ used for encrypting the message, and outputs $\genericMsg = \decrypt(\kappa, \encryptedMsg, \genericRandCoin)$. 
    
\end{itemize}
\end{defn}

\paragraph{Existing E2EE Apps \& Definition~\ref{defn:etoee}}
As mentioned, Definition~\ref{defn:etoee} does not capture all functionalities of a highly-complex modern messaging application. To the best of our knowledge, there does not exist a comprehensive definition or security analysis of a full-fledged E2EE messaging application. In fact, each E2EE application may have its own features that are different from the others so one unified definition may also not be suitable. Nonetheless, the following shows how Definition~\ref{defn:etoee} captures the functionalities provided by two existing E2EE messaging apps (that are compatible with {\sys}), namely Signal and Telegram. 

 \noindent\textbf{\underline{Signal}}
    
    \begin{itemize}
        \item \registration: During registration, Signal clients generate different types of long-term cryptographic keys and register with an identity server. The keys include "identity keys", "signed prekeys" and "one-time prekeys" \cite{cohn2020formal}. 
        
        \item \keyDerive: Signal derives all keys using key derivation functions (KDF) using the long-term keys set up during registration, and shared secrets exchanged via ratcheting. In Signal, all messages are encrypted using a unique random key \cite{cohn2020formal}. 
        
        \item \sendMessage: Signal on-wire payload has two layers of encryption. The inner layer encrypts the message using AES-CBC with the key being generated using a KDF (using shared and long-term secrets). The outer layer creates an \textit{envelope} over the encrypted message and includes message metadata that attests the sender's identity (as part of the sealed sender). This uses AES-GCM with the key derived using a random nonce shared between the sender and the recipient. The random nonce plays the role of \genericRandCoin as per definition \ref{defn:etoee}.
        
        \item \recvMessage: Given an encrypted payload, the recipient derives the keys for the two layers of encryption using shared secrets, long-term identity keys, etc., and then decrypts them one after the other. 

    \end{itemize}
    
\noindent\textbf{\underline{Telegram}}
    \begin{itemize}
        \item \registration: The registration phase in the MTProto 2.0 protocol used by Telegram corresponds to the messaging client generating a long-term, persistent shared \textit{authentication key} with the server using a key exchange protocol. The authentication key is used later for generating message keys. 

        \item \keyDerive: The key derive functionality in MTProto 2.0 generates a message-specific key using the (padded) message and a random salt. The message key is then input into a KDF along with the authentication key to derive a key that is used to encrypt the payload (padded message + salt). In the context of Telegram, the random coin \genericRandCoin corresponds to the salt used for generating the message key, which is also sent over the wire as part of the encrypted payload. 

        \item \sendMessage: The \sendMessage functionality in the context of Telegram encrypts the payload under the key derived from the message key and the authentication key using AES IGE encryption. 

        \item \recvMessage: The \recvMessage functionality receives the message key, and the authentication key and decrypts the payload. 

    \end{itemize}

\begin{defn}
\label{defn:winketoee}
A {\sys}-enabled end-to-end encrypted messaging application client, \winkEToEE makes use of a compatible end-to-end encrypted messaging application \genericEToEE, and implements the following algorithms 
$\{\registration, \keyDerive, \hidKeyDerive, \sendMessage,\\\sendHidMessage, \recvMessage, \recvHidMessage\}$


\begin{itemize}

    \item[] {\bf Functions provided by underlying E2EE app}

     \item $\registration(\secParam, \clientID)$ implements the same function as in Definition~\ref{defn:etoee}
     

      \item $\keyDerive(\secParam, \keyBundle)$ implements the same function as in Definition~\ref{defn:etoee}.
      

    \item $\sendMessage(\genericMsg, \clientID, \kappa, \genericRandCoin)$ implements the same functionality as Definition~\ref{defn:etoee} -- the random coin used in \encrypt,\genericRandCoin, is provided by {\sys}.

    
    \item $\recvMessage(\encryptedMsg, \kappa, \genericRandCoin)$ implements the same function as in Definition~\ref{defn:etoee}.


    \item[] {\bf {\sys}-specific functions}

     \item $\hidKeyDerive(\secParam)$ takes as input a security parameter, \secParam, and returns a key $\hidKey \in \{0,1\}^{\secParam}$.

    \item $\sendHidMessage(\genericMsg_1, \ldots, \genericMsg_{k+1}, \genericHidMsg, \clientID, \kappa, \hidKey)$ takes as input a set of public messages, $\genericMsg_1, \ldots, \genericMsg_{k+1}$, a hidden message \genericHidMsg, the identifier of the recipient client \clientID,  a key $\kappa$ derived from the key space of \genericEncryptionScheme using \keyDerive, and a hidden key $\kappa_H$ derived from the key space of \genericEncryptionScheme using \hidKeyDerive.
    
    It then outputs a set of encrypted messages $\{\encryptedMsg_1 = \encrypt(\kappa, \genericMsg_1, \genericRandCoin), \encryptedMsg_2 = \encrypt(\kappa, \genericMsg_2, c_1'), \ldots, \encryptedMsg_{k+1} = \encrypt(\kappa, \genericMsg_{k+1}, c_k')\}$, where $\genericRandCoin \in \{0,1\}^{\secParam}$  is a random coin (e.g., an initialization vector for a block cipher) used in the \encrypt algorithm, $c'_1 = \genericRandCoin$ and 
    $c'_2 || \ldots || c'_{k+1} = \encrypt(\hidKey, \genericHidMsg, \genericRandCoin)$.
    
    
    \item $\recvHidMessage(\encryptedMsg_1, \ldots, \encryptedMsg_{k+1}, \kappa, \hidKey, c_1', \ldots, c_{k+1}')$ takes as input a set of encrypted message $\{\encryptedMsg_1, \ldots, \encryptedMsg_{k+1}\}$, a key $\kappa$ derived from \keyDerive of \genericEToEE, a key \hidKey derived from \hidKeyDerive, and a set of encrypted hidden messages, $\{c_1', \ldots, c_{k+1}'\}$, and returns $\genericHidMsg = \decrypt(\hidKey, c_2' || \ldots || c_{k+1}', c_1')$. 

\end{itemize}
\end{defn}

\paragraph{Differences in Definitions}
There are a few critical differences between the definition of a {\sys}-enabled E2EE application (Definition~\ref{defn:winketoee}) and that of a standard E2EE messaging application (Definition~\ref{defn:etoee}).  

\begin{itemize}[nosep,leftmargin=1.6em,labelwidth=*,align=left]

    \item \sendMessage for a generic E2EE messaging application (Definition~\ref{defn:etoee}) takes as input a random coin (e.g., an IV), \genericRandCoin, which is then used in encrypting the input message. This is because instead of implementing their own pseudo-random number generators, most generic E2EE applications rely on standard PRNG implementations in external libraries or the OS. On the other hand, in a {\sys}-enabled application, the {\sys} library implements a PRNG and provides the random coin used for encryption. 

    \item \hidKeyDerive, \sendHidMessage and \recvHidMessage are available 
    \textit{only} when {\sys} is operating in public-hidden mode.

    \item When sending a hidden message with \sendHidMessage, there needs to be an appropriate number of public messages that provide opportunities to inject encrypted hidden message chunks. Thus, \sendHidMessage takes as input $k+1$ public messages. Encrypting these messages provides $k+1$ random coins for injecting hidden messages: the first random coin is used for sending the IV that is used for encrypting the hidden message, while the rest of the random coins are used for sending hidden message chunks. 
    
\end{itemize}

\subsection{Proof of Theorem~\ref{thm:winkLite:non_compromise}}
\label{sec:proof-1}

\nonCompromise*

We show that message transcripts of \genericEToEE and \winkEToEE are indistinguishable for both the public-only and public-hidden modes of operations. 

\subsubsection{Public-Only Mode}
\label{proof:public_only}
When a {\sys}-enabled E2EE messaging application, \winkEToEE, operates in public-only mode, the only difference from running the corresponding standard E2EE app is the way the \genericRandCoin randomness is generated for \sendMessage: instead of a standard PRNG library, it is provided by the {\sys} library.


As long as {\sys} properly implements random coin generation (e.g., using a PRNG), this should not impact security.

Thus, in {\em public-only mode} communication transcripts for \genericEToEE and \winkEToEE are indistinguishable if \genericEToEE uses a correctly implemented PRNG for selecting random coins. 

\subsubsection{Public-Hidden Mode}
\label{proof:public_hidden}
When \winkEToEE operates in the public-hidden mode, it replaces the random coin(s)  with semantically secure encrypted \textit{chunks} of a hidden message. 


We now show that the communication transcripts of the standard messaging application \genericEToEE, and {\sys}-enabled version of \genericEToEE operating in the public-hidden mode are indistinguishable to a PPT adversary. More precisely, if there is an adversary that can break the message privacy guarantees of a {\sys}-enabled version of \genericEToEE with non-negligible advantage, then it can also break the privacy guarantees of \genericEToEE with non-negligible advantage. 

To this end, we define the game between a challenger \ch and a PPT adversary \adv: 

\begin{enumerate}[nosep,leftmargin=1.6em,labelwidth=*,align=left]

\item \ch registers with \genericEToEE and \winkEToEE (in public-hidden mode) with \secParam and a client identifier $\clientID \in \{0,1\}^{*}$ and obtains $\keyBundle_0 = \genericEToEE.\registration(\secParam, \clientID)$ and $\keyBundle_1 = \winkEToEE.\registration(\secParam, \clientID)$ respectively.

\item \adv selects: 
\begin{itemize}
    \item a hidden message $\genericHidMsg \in \{0,1\}^{*}$ such that encrypting \genericHidMsg under the scheme \genericEncryptionScheme results in a ciphertext of length $\ell$ (in bits), and 
    \item a set of public messages $\genericMsg_1, \ldots, \genericMsg_{k+1}$ such that $\frac{\ell}{\secParam} \leq k$. \adv sends \genericHidMsg and $\genericMsg_{1}, \dots, \genericMsg_{k+1}$ to \ch. 
\end{itemize}

\item \ch flips a fair coin and selects a bit \genericBit. Then,

\begin{itemize}[nosep,leftmargin=1.6em,labelwidth=*,align=left]
    \item if $\genericBit = 0$, \ch generates the encrypted message transcripts for $\genericMsg_1, \dots, \genericMsg_{k+1}$ using $\genericEToEE.\sendMessage$, and sends the message transcripts to \adv. Specifically, \ch does the following
    
    \begin{enumerate}[nosep,leftmargin=1.6em,labelwidth=*,align=left]
        \item Computes  $\kappa_0 = \genericEToEE.\keyDerive(\secParam, \keyBundle_0)$.
        \item Computes $\{c_{0,1}, \ldots, c_{0,k+1}\}$ such that for $i \in [1, k+1]$, $c_{0,i} = \genericEToEE.\sendMessage(\genericMsg_i, \clientID, \kappa_0, \genericRandCoin_{0,i})$ where $r_{0,i} \in \{0,1\}^{\secParam}$ 
        \item Sends  $\{c_{0,1}, \ldots, c_{0,k+1}\}$ and $\{r_{0,1} \dots, r_{0,k+1}\}$ to \adv. 
    \end{enumerate}

    \item if $\genericBit = 1$, \ch generates the encrypted message transcripts for the public messages $\genericMsg_1, \dots, \genericMsg_{k+1}$ but additionally \textbf{injects the hidden message} \genericHidMsg into the transcript using 
     $\winkEToEE.\sendHidMessage$. \ch sends the message transcript to \adv. 
     Specifically, \ch does the following 
    \begin{enumerate}[nosep,leftmargin=1.6em,labelwidth=*,align=left]
        \item Computes  $\kappa_1 = \winkEToEE.\keyDerive(\secParam, \keyBundle_1)$ and $\kappa_H = \winkEToEE.\hidKeyDerive(\secParam)$.
        \item Computes $\{c_{1,1}, \ldots, c_{1,k+1}\} = \winkEToEE.\sendHidMessage(\genericMsg_1, \ldots, \genericMsg_{k+1}, \genericHidMsg, \clientID, \kappa_1, \kappa_H)$.
        \item Sends to \adv the set of encrypted messages $\{c_{1,1}, \ldots, c_{1,k+1}\}$ and $\{r_{1,1} \dots, r_{1,k+1}\}$ where $r_{1,1} \in \{0,1\}^{\secParam}$ and $r_{1,2} || \ldots || r_{1,k+1} = \encrypt(\hidKey, \genericHidMsg, r_{1,1})$.
    \end{enumerate}

\end{itemize}

\item \ch provides $\kappa_\genericBit$ to \adv on demand. \adv executes  $\recvMessage(c_{\genericBit, 1}, \kappa_\genericBit, r_{\genericBit, 1}), \dots, \recvMessage(c_{\genericBit}, \kappa_\genericBit, r_{\genericBit, k+1})$. \label{game:recv_items}

\item \adv outputs bit $\advBit$ and wins the game if $\advBit = \genericBit$.

\end{enumerate}

\paragraph{Security}
%
We now show that the adversary can win the game only with negligible advantage. 

\begin{center}
    $\prob{\advBit = \genericBit} = \frac{1}{2} + negl(\secParam)$
\end{center}

\begin{proof}

    For indistinguishability, we show the set of items received by \adv in Step \ref{game:recv_items} when $\genericBit = 0$ is indistinguishable from the items received when $\genericBit = 1$. Specifically: 

    \begin{center}
        $\{\genericMsg_1, \dots, \genericMsg_{k+1}, \genericHidMsg, \kappa_0, r_{0,1}, \dots, r_{0, k+1}, c_{0,1}, \dots, c_{0,k+1}\}$ \\ 
        $\approx_c$ \\ 
        $\{\genericMsg_1, \dots, \genericMsg_{k+1}, \genericHidMsg, \kappa_1, r_{1,1}, \dots,  r_{1, k+1}, c_{1,1}, \dots,c_{1,k+1}  \}$, 
    \end{center}

    where $\approx_c$ denotes computational indistinguishability. 

    To this end, first note that the key derivation functions \genericEToEE.\keyDerive and \winkEToEE.\keyDerive are identical by definition and the keys obtained from these functions are selected independent of messages and random coins. Hence, $\kappa_0$ and $\kappa_1$ are indistinguishable to an adversary. 

    Next, observe that $r_{0,1}, \dots, r_{0, k+1} \getsr \{0,1\}^{\secParam}$. Thus, by property of \genericEncryptionScheme, the ciphertexts $c_{0,1}, \dots, c_{0, k+1}$ are indistinguishable from random strings. Also, $r_{1,1} \getsr \{0,1\}^{\secParam}$; Thus $c_{0,1}$ and $c_{1,1}$ are indistinguishable. 
    
    Importantly however, $r_{1,2}, \dots, r_{1, k+1}$ are not randomly sampled. They are substrings of $\encrypt(\kappa_H, \genericHidMsg, r_{1,1})$ of length \secParam. 
    
    Now, assume that there is an adversary \adv' that can distinguish $c_{0,i}$ from a $c_{1,i}$ for at least one $i \in [2, k+1]$. This adversary can then also distinguish $r_{0,i}$ and $r_{1,i}$ since the only difference in parameter selection (i.e., for \encrypt when generating $c_{0,i}$, and $c_{1,i}$) is in the mechanism used for selecting the random coin. Thus \adv' can distinguish between a ciphertext produced by \genericEncryptionScheme and a random string of the same length. This violates the semantic security of \genericEncryptionScheme. And since \genericEToEE uses \genericEncryptionScheme \textit{even without {\sys}}, \adv' would then be able to break the message privacy guarantees of \genericEToEE with non-negligible advantage. 
   \end{proof}

\subsection{Proof of Theorem~\ref{thm:winkLite:messageTranscript}}
\label{sec:proof-2}

\messageTranscript*

Similarly, we show that the message transcripts generated by a {\sys}-enabled messaging application, \winkEToEE, when operating in public-hidden mode with hidden message injection, are indistinguishable from message transcripts when operating in public-only mode. We formalize using the following game between a challenger \ch and a PPT adversary \adv: 

\begin{enumerate}[nosep,leftmargin=1.6em,labelwidth=*,align=left]

\item \ch registers with \winkEToEE using \secParam and a client identifier $\clientID \in \{0,1\}^{*}$ and obtains $\keyBundle = \winkEToEE.\registration(\secParam, \clientID)$.

\item \adv selects: 
\begin{itemize}
    \item a hidden message $\genericHidMsg \in \{0,1\}^{*}$ such that encrypting \genericHidMsg under the scheme \genericEncryptionScheme results in a ciphertext of length $\ell$ bits, and 
    \item a set of public messages $\genericMsg_1, \ldots, \genericMsg_{k+1}$ such that $\frac{\ell}{\secParam} \leq k$. 
\end{itemize}

\item \adv sends \genericHidMsg and $\genericMsg_{1}, \dots, \genericMsg_{k+1}$ to \ch. 

\item \ch computes a key $\kappa = \winkEToEE.\keyDerive(\secParam, \keyBundle)$. 

\item \ch flips a fair coin and selects a bit \genericBit. Then, 

\begin{itemize}[nosep,leftmargin=1.6em,labelwidth=*,align=left]
    \item if $\genericBit = 0$, \ch generates the encrypted message transcript for the public messages $\genericMsg_1, \dots, \genericMsg_{k+1}$ using  $\winkEToEE.\sendMessage$, and sends the message transcript to \adv. Specifically, \ch does the following 
    
    \begin{enumerate}[nosep,leftmargin=1.6em,labelwidth=*,align=left]
        \item Computes $\{c_{0,1}, \ldots, c_{0,k+1}\}$ such that for $i \in [1, k+1]$, $c_{0,i} = \winkEToEE.\sendMessage(\genericMsg_i, \clientID, \kappa, \genericRandCoin_{0,i})$ where $r_{0,i} \in \{0,1\}^{\secParam}$ 
        \item Sends  $\{c_{0,1}, \ldots, c_{0,k+1}\}$ and $\{r_{0,1} \dots, r_{0,k+1}\}$ to \adv. 
    \end{enumerate}

    \item If $\genericBit = 1$, ch generates the encrypted message transcripts for the public messages $\genericMsg_1, \dots, \genericMsg_{k+1}$ but additionally \textbf{injects the hidden message} \genericHidMsg into the transcript using 
     $\winkEToEE.\sendHidMessage$. \ch sends the message transcript to \adv. Specifically, \ch does the following 
    \begin{enumerate}[nosep,leftmargin=1.6em,labelwidth=*,align=left]
        \item Computes  $\kappa_H = \winkEToEE.\hidKeyDerive(\secParam)$.
        \item Computes $\{c_{1,1}, \ldots, c_{1,k+1}\} = \winkEToEE.\sendHidMessage(\genericMsg_1, \ldots, \genericMsg_{k+1}, \genericHidMsg, \clientID, \kappa, \kappa_H)$.
        \item Sends to \adv the set of encrypted messages $\{c_{1,1}, \ldots, c_{1,k+1}\}$ and $\{r_{1,1} \dots, r_{1,k+1}\}$ where $r_{1,1} \in \{0,1\}^{\secParam}$ and $r_{1,2} || \ldots || r_{1,k+1} = \encrypt(\hidKey, \genericHidMsg, r_{1,1})$.
    \end{enumerate}

\end{itemize}

\item \ch provides $\kappa$ to \adv on demand. 
\item \adv computes  $\recvMessage(c_{\genericBit, 1}, \kappa, r_{\genericBit, 1}), \dots, \recvMessage(c_{\genericBit}, \kappa, r_{\genericBit, k+1})$.

\item \adv outputs bit $\advBit$ and wins the game if $\advBit = \genericBit$.
\end{enumerate}

\paragraph{Security}
%
We now show that the adversary can win the game only with negligible advantage. 

\begin{center}
    $\prob{\advBit = \genericBit} = \frac{1}{2} + negl(\secParam)$
\end{center}

\begin{proof}
    This follows from two results in the preceding sections.
    
    Firstly, the message transcripts for the public-only mode of operation are indistinguishable from the transcripts for the standard messaging application, \genericEToEE, if {\sys} selects random coins using a correct PRNG for \sendMessage (Section~\ref{proof:public_only}). 
    
    Secondly, the message transcripts of {\sys} operating in public-hidden mode are indistinguishable from the message transcripts of \genericEToEE (in proof of Theorem~\ref{thm:winkLite:non_compromise}). 
    
    It straightforwardly follows that the message transcripts of the public-only mode of operation of \winkEToEE are indistinguishable from the message transcripts of \winkEToEE operating in the public-hidden mode of operation if the encryption used in \sendMessage and \sendHidMessage is $IND\$-CPA$ secure and produces ciphertexts indistinguishable from random. 
    
    We omit further details of this straightforward reduction.  
\end{proof}


%

\subsection{Proof of Theorem \ref{thm:winkLite:executionTranscript}}
\label{sec:proof-3}

\executionTranscript*

\begin{proofsketch}
First, note that the mode of operation in {\sys} is selected based on a secret user PIN stored in the Secure World, not available to the adversary. 

In public-only mode, the Normal World adversary observes the public message inputs in plaintext. This is followed by a context switch to the Secure World for encryption, and the Normal World subsequently obtains the ciphertexts and the related random coins (e.g., IV), and sends it across the wire. Similarly, after receiving an encrypted public message (and its related random coins), the Normal World adversary observes a context switch to the Secure World for decryption and subsequently observes the public message plaintext.

In public-hidden mode, in addition to the public message, the user inputs a hidden message directly into the secure World using the Secure I/O interface not accessible to the adversary. The hidden message is encrypted with IND-CPA secure encryption, and the ciphertext is used as a random coin for public message encryption. Similarly, a hidden message received with an encrypted public message is decrypted and directly output to the user in the Secure World. IND-CPA security ensures that the adversary cannot distinguish between random coins and encrypted hidden messages. 

Additionally, hidden message inputs/outputs are initiated when Wink receives in Secure World the Wink-specific hardware interrupt triggered upon a user button press. Thus, timing Secure World execution does not provide any advantage to an adversary incapable of monitoring hardware interrupts in the Normal World. Also, the adversary does not gain any information by measuring the time period e.g., using the system's clock, between two public messages since the difference between the time when a user sends/receives a (public) message and sends/replies with the next message is typically arbitrary. Note that this time window may also be used for hidden message inputs. 

Public-only and public-hidden messaging encryption/decryption requests are also indistinguishable in terms of processing time from outside the TEE. The injection of hidden messaging only implies replacing generated random IVs with hidden messages already encrypted during the processing of \sys{} hardware interrupts described above.
\end{proofsketch}

\subsection{Proof of Theorem \ref{thm:winkHeavy}}
\label{sec:proof-4}

\winkHeavy*

\begin{proofsketch}
By design, the PIN for determining the mode of operation in {\sys} is never accessible to the Normal World, including the kernel. The TrustZone-enforced hardware isolation prevents the REE kernel from observing Secure World operations. As a result, the REE kernel can only detect and time TEE entry and exists, which may occur either due to hardware interrupts or SMCs. 

The utilization of a unified Secure World UI for both public and hidden messaging eliminates the need for dedicated hardware interrupts for hidden messaging. Instead, the hidden messaging interface is always shown alongside the one used for public messages \textit{when the device operates in public-hidden mode}. Thus, REE kernel adversaries are limited to only tracking utilization of the unified interface which might also include hidden message injection. 

It is infeasible for the REE kernel to determine when the Secure World UI is used for hidden message I/O, in addition to the public message I/O since a user can spend an arbitrary amount of time writing or reading public messages. Therefore, the time spent on public + hidden I/O can be plausibly attributed 
to public-only I/O. Further, the time required for performing cryptographic operations on hidden messages is insignificant in comparison to (public) user I/O and cannot be used for distinguishing the mode of operation with more than negligible over-guessing. 
%
%
\end{proofsketch}

\end{document}